\documentclass[11pt]{article}
\usepackage{amssymb,amsmath,amsthm}
\usepackage{graphicx}
\begin{document}

\title{\bfseries On the Capacity of the Heat Channel, Waterfilling in the Time-Frequency Plane, and a 
C-NODE Relationship\footnote{The material in this paper was presented in part at the IEEE International 
Symposium on Information Theory, Seoul, Korea, June 28--July 3, 2009 \cite{Ham2009}.}}

\author{Edwin Hammerich\\
    \small Ministry of Defence\\
    \small Kulmbacher Str. 58--60, D-95030 Hof, Germany\\
    \small E-mail: edwin.hammerich@ieee.org}
\date{}
\maketitle

\begin{abstract}
The heat channel is defined by a linear time-varying (LTV) filter with additive white Gaussian 
noise (AWGN) at the filter output. The continuous-time LTV filter is related to the heat kernel 
of the quantum mechanical harmonic 
oscillator, so the name of the channel. The channel's capacity is given in closed form 
by means of the Lambert W function. Also a waterfilling theorem in the time-frequency plane for the 
capacity is derived. It relies on a specific Szeg\H{o} theorem for which an essentially self-contained 
proof is provided. Similarly, the rate distortion function for a related nonstationary source is given 
in closed form and a (reverse) waterfilling theorem in the time-frequency plane is derived. Finally, a 
second closed-form expression for the capacity of the heat channel based on the detected perturbed 
filter output signals is presented. In this context, a precise differential 
connection between channel capacity and the normalized optimal detection error (NODE) is revealed. This 
C-NODE relationship is compared with the well-known I-MMSE relationship connecting mutual information with 
the minimum mean-square error (MMSE) of estimation theory.
\end{abstract}

\theoremstyle{definition} % benoetigt package amsthm
\newtheorem{definition}{Definition}
\newtheorem{theorem}{Theorem}
\newtheorem{proposition}{Proposition}
\newtheorem{lemma}{Lemma}
\newtheorem{remark}{Remark}

\newcommand{\dxdxi}{\,\mathrm{d}x\,\mathrm{d}\xi}
\newcommand{\dtdomega}{\,\mathrm{d}t\,\mathrm{d}\omega}
\newcommand{\domega}{\,\mathrm{d}\omega}
\newcommand{\df}{\,\mathrm{d}f}
\newcommand{\dt}{\,\mathrm{d}t}
\newcommand{\dx}{\,\mathrm{d}x}
\newcommand{\dy}{\,\mathrm{d}y}
\newcommand{\dnachdsnr}{\frac{\mathrm{d}}{\mathrm{d}\,\snr} }
\newcommand{\mmse}{\mathrm{mmse}}
\newcommand{\snr}{\mathrm{snr}}
\newcommand{\node}{\mathrm{node}}

\section{Introduction}\label{Section_I}
\begin{figure}
\centering
\includegraphics[width=3.5in]{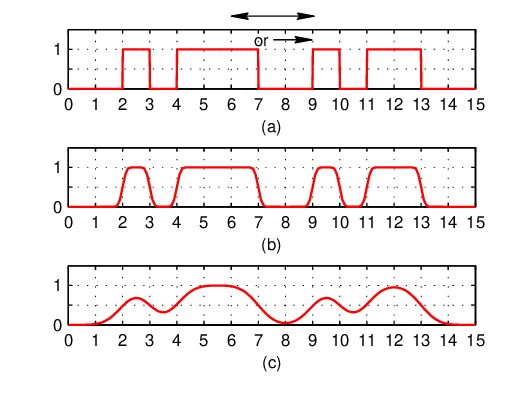}
\caption{Spreading of heat/signal spread because of dispersion in an optical fiber (attenuation not 
regarded). (a)~Initial temperature distribution/fiber input signal. (b)~Temperature, shortly
after/output signal, short fiber. 
%($\beta=10$). 
(c)~Temperature, later time/output signal, longer fiber.
%($\beta=2$). 
Arrows indicate direction of spatial propagation of heat ($\leftrightarrow$) or optical intensity 
($\rightarrow$).}
\label{Figure_1}
\end{figure}

The conduction of heat in solid bodies was mathematically described and solved by 
Joseph Fourier in his fundamental 1822 treatise \textit{Th\'{e}orie analytique de la chaleur} 
\cite{Fourier}. In one dimension, e.g., in case of a heat-conducting insulated wire, his
description results in the partial differential equation (known as heat equation)
\[
   \frac{\partial u}{\partial t}=k\frac{\partial^2u}{\partial x^2},
\]
in which $u=u(x,t)$ is temperature at time $t\ge0$ at any point $x$ and $k$ is a positive constant 
depending on the material. Given the initial temperature distribution $f(x)$ for a wire of infinite 
length, Fourier's solution to the heat equation is% \cite[Articles 354, 370]{Fourier}
\begin{equation}
   u(x,t)=\frac{1}{\sqrt{4\pi kt}}\int_{-\infty}^\infty e^{-\frac{(x-y)^2}{4kt}}f(y)\dy.\label{solheat}
\end{equation}
Since, in general, for any time $t>0$ the inversion of the integral transform appearing in 
(\ref{solheat}) is unfeasible in practice \cite{TA}, we observe an unavoidable loss of ``information" 
(in a preliminary, informal sense). 
%with the propagation of heat.
In Fig.~\ref{Figure_1}, several temperature distributions $u(x,t)$ are depicted 
showing how the initial one is gradually smeared out by the propagation of heat. 

\begin{figure}
\centering
\includegraphics[width=3.5in]{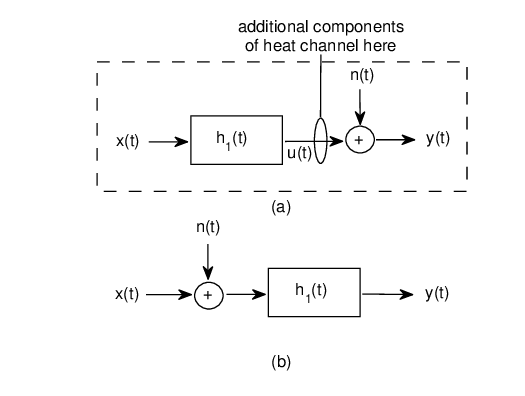}
\caption{Continuous-time channels. (a)~Gallager model \cite[Fig.~8.4.1]{Gallager} (in dashed box) with
LTI filter with impulse response $h_1(t)$. 
(b)~Bandlimited Gaussian Channel; here the LTI filter is an ideal low-pass filter. Input 
signal $x(t)$ always power-limited; noise signal $n(t)$ realization of white Gaussian noise.}
\label{Figure_2}
\end{figure}

A similar situation, in principle known since the earliest days of cable communication \cite{Falconer}, 
arises in fiber optics. In a transmission through a (single-mode) optical fiber, signals experience 
besides attenuation a spread over time due to (chromatic) dispersion (see, e.g., \cite{Agrawal}, 
\cite{EKWetal}). A frequently used model for dispersion in an optical fiber \cite{Agrawal} is a linear 
time-invariant (LTI) filter with impulse response (cf. Fig.~\ref{Figure_2})
\begin{equation}
   h_1(t)=\frac{1}{\sqrt{2\pi}(1/\beta)}e^{-\frac{1}{2}\frac{t^2}{(1/\beta)^2}},  \label{h1}
\end{equation}
i.e., a Gaussian filter with the standard deviation $1/\beta$, $\beta\in(0,\infty)$ 
characterizing dispersion (the parametrization is chosen to fit later notation). The fiber input/output 
relation is now given by the convolution integral
\begin{equation}
   u(t)=a\int_{-\infty}^\infty h_1(t-t')f(t')\dt', \label{LTI}
\end{equation}
where $f\in L^2(\mathbb{R})$ is the finite-energy input signal and $u(t)$ the output, the constant 
factor $a\in (0,1]$ representing attenuation. Except for the factor $a$ and change of physical 
dimension from position (variable $x$) to time (variable $t$), we now observe perfect analogy between 
Eqs. (\ref{solheat}) and (\ref{LTI}). As a consequence, in Fig.~\ref{Figure_1} also the degradation 
of an optical signal---initially a sequence of bits (here, binary symbols) obtained by intensity 
modulation and on-off keying---by dispersion in an optical fiber is displayed. Obviously, dispersion 
limits the information throughput of an optical fiber because of intersymbol interference (ISI). A 
maximum attainable bit rate can be estimated by considering as fiber input a sequence of unit impulses 
separated by time intervals of duration $\Delta t$, the output then being a sequence of Gaussian pulses of 
standard deviation $\tau=1/\beta$. In order to cope with ISI, a popular criterion is 
$\tau\le\Delta t/4$ \cite{Agrawal} resulting in our case for the bit rate $R_b=1/{\Delta t}$ 
(in binary symbols per second) in the estimate
\begin{equation}
   R_b\le \frac{\beta}{4}.  \label{RuleOfThumb}
\end{equation}

If the fiber output signal is corrupted by noise, this rule of thumb becomes questionable. In case of 
AWGN, typically caused by an optical amplifier \cite{Agrawal}, 
\cite{WE}, we arrive at a continuous-time (or waveform) channel following the model in Gallager's 
book~\cite{Gallager}; see Fig.~\ref{Figure_2}(a). Here, of course, we supposed that, 
e.g., the power spectral density (PSD) of the AWGN is independent of the input (see \cite{Moser} for an 
opposite situation), and the input power is not so high that 
arising nonlinearities \cite{Agrawal} would destroy the linear model (\ref{LTI}); we refer to 
\cite{EKWetal} for a variety of possible other perturbations that limit the capacity of optical fiber 
communication systems. 

In this paper we investigate the LTV filter (or operator) 
$\boldsymbol{P}_\delta^{(\gamma)}:L^2(\mathbb{R})\rightarrow L^2(\mathbb{R})$ given by
\begin{align}
(\boldsymbol{P}_\delta^{(\gamma)}f)(t)&=e^{-\frac{t^2}{2\alpha^2}}            \label{TFLO}\\
    &\cdot\frac{\beta}{\sqrt{2\pi\cosh\delta}}\int_{-\infty}^\infty \exp\left[-\frac{\beta^2}{2}
	                  \left(\frac{t}{\cosh\delta}-t'\right)^2\right]f(t')\dt',  \nonumber
\end{align}
where $t$ is time and $\alpha,\,\beta$ are any positive numbers satisfying $\alpha\beta>1$ and the 
positive parameters $\gamma,\,\delta$ are defined by $\gamma^2=\alpha/\beta$, $\coth\delta=\alpha\beta$. 
This operator, in its original form introduced as time-frequency localization operator in signal 
analysis \cite{Daubechies2}, was used in the present form as prefilter in a sampling 
theorem \cite{Ham2004}. The Fourier transform of the filter output signal 
$g=\boldsymbol{P}_\delta^{(\gamma)}f$ is
\begin{align}
\hat{g}(\omega)&=e^{-\frac{\omega^2}{2\beta^2}}                                  \label{TFLO_freq}\\
	&\cdot\frac{\alpha}{\sqrt{2\pi\cosh\delta}}\int_{-\infty}^\infty 
         \exp\left[-\frac{\alpha^2}{2}\left(\frac{\omega}{\cosh\delta}-\omega'\right)^2\right]
		                                         \hat{f}(\omega')\domega',        \nonumber
\end{align}
where $\omega$ is angular frequency and for the Fourier transform the convention 
$\hat{f}(\omega)=(2\pi)^{-1/2}\int_{-\infty}^\infty e^{-it\omega}f(t)\dt$ has been used 
\cite{Daubechies1}. The condition $\alpha\beta>1$ is now seen as imposed by the uncertainty principle of 
communications~\cite{Gabor}. Interestingly enough, the kernel of operator 
$\boldsymbol{P}_\delta^{(\gamma)}$ coincides 
with the heat kernel \cite{Getzler} of the quantum mechanical harmonic 
oscillator.\footnote{In \cite[p. 114]{Getzler}, the heat kernel of the one dimensional quantum 
mechanical harmonic oscillator with Hamiltonian $H=-\frac{d^2}{dx^2}+a^2x^2$ takes the form of the 
kernel of operator (\ref{TFLO}) after the substitution $2at=\delta,\,a=\gamma^{-2}$.} 
Because of the Gaussian prefactor on the right-hand side (RHS) of Eq.~\eqref{TFLO_freq}, the Fourier 
transform $\hat{g}(\omega)$ of the filter output signal decays exponentially outside of the interval 
$[-\pi\beta,\pi\beta]$ [provided that the energy of the input signal $f(t)$ is not too high]. Thus, 
$g(t)$ may be considered an approximately bandlimited signal of approximate bandwidth $W=\beta/2$ in 
positive frequencies measured in hertz.

If $u(t)$ is the output signal of the LTI filter (\ref{LTI}) (where  $a=1$) with impulse 
response (\ref{h1}) upon input signal $f(t)$, then the output signal $g(t)$ of the LTV 
filter~(\ref{TFLO}) may be written as
\begin{equation}
  g(t)=e^{-\frac{t^2}{2\alpha^2}}\cdot(\cosh\delta)^{-\frac{1}{2}}u(t/\cosh\delta). \label{IOheat}
\end{equation}
Thus, $g(t)$ is just the dilated LTI filter output signal $u(t)$ multiplied with a Gaussian time window. 
In Fig.~\ref{Figure_2}(a), those two operations are quoted as additional components of the heat channel, 
the latter meaning the continuous-time LTV channel formed by the LTV filter (\ref{TFLO}) and subsequent 
AWGN. As a consequence, the heat channel can be used to lower bound the capacity of an optical fiber in 
the presence of dispersion and AWGN. It will also allow us to design short-time pulses attaining that 
bound.

\section{The Heat Channel}\label{Section_II}
\begin{definition}\label{DefHC}
The \emph{heat channel} is the continuous-time LTV channel
\begin{equation}
   \tilde{g}(t)=(\boldsymbol{P}_\delta^{(\gamma)}f)(t)+n(t),\,-\infty<t<\infty,  \label{HC}
\end{equation}
where $\boldsymbol{P}_\delta^{(\gamma)}$ is the LTV filter \eqref{TFLO}, the real-valued filter input 
signals $f(t)$ are of finite energy and the noise signals  $n(t)$ at the filter output are realizations 
of white Gaussian noise with two-sided PSD $N_0/2=\theta^2>0$.
\end{definition}

Henceforth, we use the parameter 
\begin{equation}
   \rho=e^{-\delta},\,\delta=\mathrm{arccoth}(\alpha\beta)=\frac{1}{\alpha\beta}+
                       \frac{1}{3(\alpha\beta)^3}+\ldots\,. \label{def_rho}
\end{equation}
Note that $\lim_{\alpha\beta\rightarrow\infty}\frac{\delta(\alpha\beta)}{(\alpha\beta)^{-1}}=1$ (or 
$\delta\sim\frac{1}{\alpha\beta}$ as $\alpha\beta\rightarrow\infty$ for short). We shall now reduce the 
continuous-time heat channel to a (discrete) vector Gaussian channel following the approach in 
\cite{Gallager} for (LTI) waveform channels.

\subsection{Diagonalization of the Filter}
As shown in \cite{Daubechies2} in the radial case $\alpha=\beta$, i.e., $\gamma=1$ (see 
\cite{Ham2004} for the general case $\gamma>0$), the operator $\boldsymbol{P}_\delta^{(\gamma)}$ in 
(\ref{TFLO}) has eigenvalues $\rho^{k+\frac{1}{2}},\,k=0,1,\ldots,$ with corresponding eigenfunctions 
\[
  (D_\gamma \psi_k)(t)=\gamma^{-\frac{1}{2}}\psi_k(t/\gamma),
\]
where $\psi_k(t)=(2^k k!\sqrt{\pi})^{-1/2}H_k(t) e^{-t^2/2}$ is the $k$th Hermite function, 
$H_k(t)=e^{t^2}(-d/dt)^k e^{-t^2}$ being the $k$th Hermite polynomial \cite{Abramowitz}.
\begin{figure}
\centering
\includegraphics[width=3.5in]{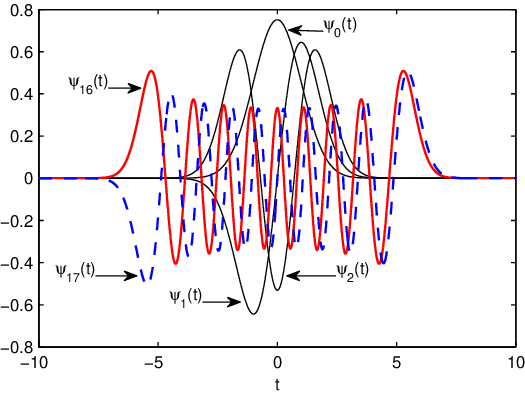}
\caption{Hermite functions $\psi_0,\,\psi_1,\,\psi_2,\,\psi_{16},\,\psi_{17}$; 
$\psi_0$ is a normalized Gaussian function. Strong decay in time is complemented by the same 
behaviour in the frequency domain since Hermite functions are eigenfunctions of the Fourier transform.}
\label{Figure_3}
\end{figure}
Since $\{D_\gamma \psi_k;\,k=0,1,\dots\}$ forms a complete orthonormal basis of 
$L^2(\mathbb{R})$, any function $f\in L^2(\mathbb{R})$ has an expansion 
$f(t)=\sum_{k=0}^\infty x_k\,(D_\gamma \psi_k)(t)$ where the coefficient sequence $x_0,x_1,\ldots$ is an 
element of the space $\ell^2(\mathbb{N}_0)$ of square-summable complex sequences with index set 
$\mathbb{N}_0=\{0,1,\ldots\}$. Hence for any filter input signal $f\in L^2(\mathbb{R})$, the filter 
output signal has the representation
\begin{equation}
  (\boldsymbol{P}_\delta^{(\gamma)}f)(t)=
                      \sum_{k=0}^\infty\rho^{k+\frac{1}{2}}x_k\,(D_\gamma \psi_k)(t), \label{EXP}
\end{equation}
where $x_k=\langle f,D_\gamma \psi_k\rangle$, 
$\langle f_1,f_2 \rangle=\int_{-\infty}^\infty f_1(t)\overline{f_2(t)}\dt$ denoting the inner product 
in $L^2(\mathbb{R})$. The new coefficient sequence is $(\rho^{k+\frac{1}{2}}x_k)_{k=0}^\infty$ and again 
an element of $\ell^2(\mathbb{N}_0)$. Thus, the filter $\boldsymbol{P}_\delta^{(\gamma)}$ is reduced to 
a diagonal linear transformation in $\ell^2(\mathbb{N}_0)$.

In Fig.~\ref{Figure_3}, some Hermite functions are depicted; observe their strong decay in time (and 
frequency). 

\subsection{Discretization of the Heat Channel}\label{Section_II-B}
The perturbed filter output signal is $\tilde{g}(t)=g(t)+n(t)$, where the noise signal $n(t)$ is as 
described in Def.~\ref{DefHC}. To extract as much information as possible from $\tilde{g}(t)$ we apply 
\textit{optimal detection} (due to North \cite{North}), for example by means of a bank of matched 
filters \cite{Turin}, in our case LTI filters with impulse response $h_k(t)=(D_\gamma\psi_k)(-t)$. 
When applied to the noisy signal $\tilde{g}(t)$ as input, we get for the matched filter output signals 
sampled at time zero
\[
  \int_{-\infty}^\infty h_k(0-t')\tilde{g}(t')\dt'
                             =\rho^{k+\frac{1}{2}}x_k+\int_{-\infty}^\infty h_k(-t')n(t')\dt'.
\]
From the theory of LTI filters we know that the integral 
on the RHS evaluates to a realization $n_k$ of a zero-mean Gaussian random variable 
$N_k$ with the variance $\theta^2\int_{-\infty}^\infty h_k^2(-t)\dt$. Since any waveform 
$D_\gamma\psi_k$ has $L^2$ norm one, the variance of $N_k$ is $\theta^2$ and, thus, does not depend on 
$k$. Moreover, because of orthogonality of the waveforms, the random variables $N_k$ are independent. 
Consequently, the detection errors $n_k$ are realizations of independent identically distributed 
($\mathrm{i.i.d.}$) zero-mean Gaussian random variables $N_k\sim\mathcal{N}(0,\theta^2),k=0,1,\ldots$. 
Note that the noise PSD $\theta^2$, measured in watts/Hz, has also the dimension of an energy.

Instead of $y_k=\rho^{k+\frac{1}{2}}x_k$ we now have obtained $\hat{y}_k=y_k+n_k$. So we get the 
estimate $\hat{x}_k=\rho^{-k-\frac{1}{2}}\hat{y}_k=x_k+z_k$ for $x_k$, where $z_k$ are realizations of 
independent Gaussian random variables $Z_k\sim\mathcal{N}(0,\theta^2\rho^{-2k-1}),\,k=0,1,\ldots$. Thus, 
we are led to the infinite-dimensional vector Gaussian channel
\begin{equation}
  Y_k=X_k+Z_k,\,k=0,1,\ldots,  \label{discr_HC}
\end{equation}
where the noise $Z_k$ is distributed as described.

\begin{figure}
\centering
\includegraphics[width=3.5in]{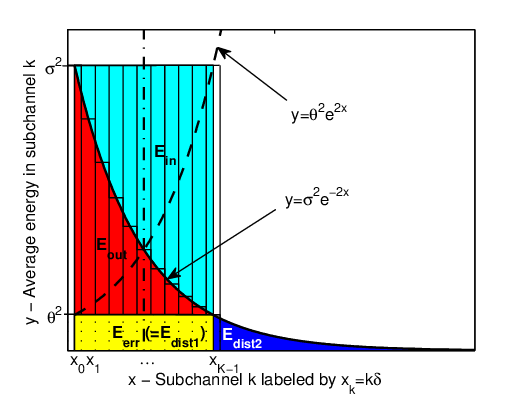}
\caption{Balance of (average) energies around the heat channel: input/output energy 
($E_\mathrm{in}/E_\mathrm{out}$), energy of measurement error ($E_\mathrm{err}$), and distortion 
($E_\mathrm{dist1}+E_\mathrm{dist2}$). Subchannels displayed at distance 
$\delta\sim\frac{1}{\alpha\beta}$ apart as $\alpha\beta\rightarrow\infty$.}
\label{Figure_4}
\end{figure}

$\boldsymbol{X}^K=(X_0,\ldots,X_{K-1})^T$ will denote a $K$-dimensional column vector, $K\in\mathbb{N}$, 
of not necessarily independent random variables $X_k$. For any average input energy $S>0$, the vector 
Gaussian channel consisting of the first $K$ subchannels of the heat channel has 
capacity~\cite{Shannon1948}
\begin{equation}
  C^K(S)=\max_{\mathbb{E}\|\boldsymbol{X}^K\|^2\le S} I(\boldsymbol{X}^K;\boldsymbol{Y}^K),
                                                                                \label{capacity}
\end{equation}
where $I(\boldsymbol{X}^K;\boldsymbol{Y}^K)$ is the mutual information between random input vector 
$\boldsymbol{X}^K$ and corresponding random output vector $\boldsymbol{Y}^K$, $\boldsymbol{X}^K$ subject 
to the average energy constraint $\mathbb{E}\|\boldsymbol{X}^K\|^2=\sum_{k=0}^{K-1}\mathbb{E}X_k^2\le S$. 
The noise variances $\theta^2\rho^{-2k-1},\,k=0,1,\ldots,$ are monotonically increasing and unbounded. 
Consequently, by reason of the well-known waterfilling argument~\cite{Cover}, for any fixed average 
input energy $S$ the sequence of capacities $C^1(S),C^2(S),\ldots$ eventually becomes constant. We 
define
\begin{equation}
  C(S)=\lim_{K\rightarrow\infty}C^K(S) \label{def_cap}
\end{equation}
as the capacity of the heat channel; $C(S)$ is measured in bits per channel use (or transmission). 
In Fig.~\ref{Figure_4}, $E_\mathrm{in}=S$, the $K$ subchannels to be read in reverse order (as will 
become clear in Section~\ref{Section_III-A}); further details of Fig.~\ref{Figure_4} will be described 
in the text.

\subsection{Example: Dispersion and Amplifier Noise in Fiber~Optics}\label{Section_II-C}
When, as supposed here, in the fiber-optic transmission intensity modulation is applied, real-valued 
input signals $f\in L^2(\mathbb{R})$ need to be replaced by waveforms $\tau_0+f(t)$ where $\tau_0$ is a 
fixed positive number chosen large enough so that the resulting signals are nonnegative with high 
probability [cf. Fig.~\ref{Figure_9}(a), below]. Dispersion is modeled through an LTI filter with 
impulse response as in (\ref{h1}). The dispersion parameter $\beta=\beta(T)\in(0,\infty)$ in (\ref{h1}) 
depends on transmission time $T$ as well as the coefficient $a=a(T)\in(0,1]$ in (\ref{LTI}) representing 
attenuation. Then, the fiber output signal is the waveform $a\tau_0+u(t)$ where $u(t)$ is given by 
Eq.~(\ref{LTI}). 

In order to create a heat channel, choose at the receiver any fixed time parameter $\alpha$ with the 
property that $\alpha>1/\beta$. When the product $\alpha\beta$ is large, the dilation in (\ref{IOheat}) 
may be neglected in practice (for simplicity of exposition we imagine of having performed the dilation). 
Now, apply a variable density neutral filter of appropriate characteristic (an optical device, 
see \cite{Gross}) to effect---in the spatial domain---a multiplication of the signal with the time 
window $w(t)=\exp[-t^2/(2\alpha^2)]$ as shown in (\ref{IOheat}). Next, amplify the obtained signal by 
factor $1/a$. When an optical amplifier is employed, the resulting signal is $w_1(t)+g(t)+n(t)$ 
where $w_1(t)\propto w(t)$, $g=\boldsymbol{P}_\delta^{(\gamma)}f$, and $n(t)$ is a 
realization of white Gaussian noise (properly modeling the impairment of an optical signal by an 
optical amplifier; see \cite{WE}). After opto electric conversion (possibly adding anew white Gaussian 
noise to the signal), remove the known signal component $w_1(t)$. Finally, use as detection 
device a bank of matched filters with impulse responses $h_k(t),\,k=0,\ldots,K-1,$ as given in 
Section~\ref{Section_II-B}; $K$ is a known number depending on the average input energy $S$. Thus, 
we have implemented a heat channel in an optical fiber communication system.

\subsection{Degrees of Freedom of Filter Output Signals}\label{Section_II-D}
Here, we give an explanation for the time-frequency product $\alpha\beta$ that will occur very frequently 
in the sequel.

The Wigner-Ville spectrum (WVS) of the response of filter $\boldsymbol{P}_\delta^{(\gamma)}$ on white 
Gaussian noise (see Appendix A for details) is the bivariate function
\begin{equation}
  \boldsymbol{\Phi}(t,\omega)=\frac{\sigma^2}{2\pi}\cdot\frac{1}{\cosh\delta}
                  \exp\left(-\frac{t^2}{\alpha^2}-\frac{\omega^2}{\beta^2}\right). \label{WVS}
\end{equation}
In general, the WVS of a nonstationary stochastic process gives its density of (average) energy in the 
time-frequency plane; see, e.g.,  \cite{Flandrin}, \cite{Hlawatsch}. Consequently, in our case, 
the energy of filter output signals would occupy an ellipse-shaped region in the time-frequency plane 
with unsharp boundary. Regarding the WVS (\ref{WVS}) (after a normalization) as a bivariate Gaussian 
probability density function, we describe this region by an approximation known as \emph{ellipse of 
concentration} (EOC) in probability theory \cite{Cramer}.
\begin{figure}
\centering
\includegraphics[width=3.5in]{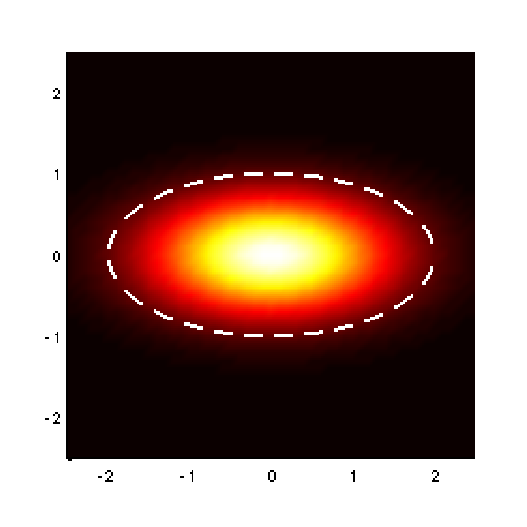}
\caption{Energy density and EOC (dashed line) for white Gaussian noise response of 
filter (\ref{TFLO}) in the limiting case $\alpha\rightarrow\sqrt{2}+,\,\beta\rightarrow(1/\sqrt{2})+$; 
graphical appearance typical also for other (admissible values) of $\alpha,\beta$.}
\label{Figure_5}
\end{figure}
We obtain as EOC the region 
$\mathcal{A}_\mathrm{c}=\{(t,\omega)\in\mathbb{R}^2;t^2/\alpha^2+\omega^2/\beta^2\le 2\}$ with area 
$A_c=2\pi\alpha\beta$ \cite{Ham2004}. As reported in \cite[p. 23]{Daubechies1}, in physics a region 
in phase space (or time-frequency plane such as here) with area $A$ corresponds to $A/(2\pi)$ 
``independent states" (when $A$ is sufficiently large). Since here $A_\mathrm{c}/(2\pi)=\alpha\beta$, 
the time-frequency product $\alpha\beta$ would describe the ``dimension" of the filter output space 
(when $\alpha\beta$ is sufficiently large), i.e., the degrees of freedom (DOFs) of filter output 
signals. In Fig.~\ref{Figure_5}, the energy density in the time-frequency plane as given by the 
WVS (\ref{WVS}) is illustrated.

\section{Channel Capacity in Terms of Channel Input, and Optimal Signaling}\label{Section_III}
In this section we derive a closed formula for the capacity of the heat channel in terms of average 
energy of the channel input signal along with a method of capacity achieving (optimal) signaling. A 
characterization of channel capacity by waterfilling in the time-frequency plane is also given. The 
following definition proves to be useful.
\begin{definition} \label{def_2} For any two functions $A,\,B:(1,\infty)\rightarrow\mathbb{R}$ the 
notation $A\doteq B$ means 
\[
   \lim_{x\rightarrow\infty}\frac{A(x)-B(x)}{x}=0,
\]
or, equivalently, $A(x)=B(x)+o(x)$ as $x\rightarrow\infty$.\footnote{We use the standard Landau symbols 
little-o, $o(\cdot)$, and big-O, $O(\cdot)$.}
\end{definition}

In our context, $x$ will always be the time-frequency product $\alpha\beta>1$. Thus, $A\doteq B$ implies 
that $A(\alpha\beta)/(\alpha\beta)=B(\alpha\beta)/(\alpha\beta)+\epsilon$ where $\epsilon\rightarrow 0$ 
as $\alpha\beta\rightarrow\infty$.

\subsection{Channel Capacity in Closed Form}\label{Section_III-A}
The function $y=w_0(x),\,x\ge 0,$ occurring in the next theorem is the inverse function of 
$y=(2x-1)e^{2x}+1,\,x\ge 0$ (see also Fig.~\ref{Figure_6}).
\begin{theorem}\label{form_1} Assume that the average energy $S$ of the input signal depends on 
$\alpha\beta$ such that $S(\alpha\beta)=O(\alpha\beta)$ as $\alpha\beta\rightarrow\infty$. 
Then for the capacity (in bits per transmission) of the heat channel it holds
\begin{equation}
  C(S)\doteq\frac{\alpha\beta}{2}\left[w_0\left(\frac{S}{(\alpha\beta/2)\theta^2}\right)
                                                               \right]^2\log_2 e.\label{C1b}
\end{equation}		  
\end{theorem}
\begin{proof}[Proof:] The proof is accomplished by waterfilling \cite[Th.~7.5.1]{Gallager}, \cite{Cover}. Let 
$\nu_k^2=\theta^2\rho^{-2k-1},\,k=0,1,\ldots,$ be the variance of 
noise in the $k$th subchannel of the discetized heat channel~\eqref{discr_HC}. The positive number 
$\sigma$ is defined by the condition
\begin{equation}
  S = \sum_{k=0}^\infty (\sigma^2-\nu_k^2)^+=\sum_{k=0}^{K-1} (\sigma^2-\nu_k^2),\label{def_sigma}
\end{equation}
where $x^+\triangleq\max\{0,x\},\,x\in\mathbb{R},$ and $K=\max\{k\in\mathbb{N};\nu_{k-1}^2<\sigma^2\}$ 
is the number of subchannels in the resulting finite-dimensional vector Gaussian channel. With 
increasing time-frequency product $\alpha\beta$, $\delta=\delta(\alpha\beta)$ (now acting as increment) 
tends to 0 so that
\begin{eqnarray}
  S\cdot\delta &=& \sum_{k=0}^{K-1} (\sigma^2-\theta^2 e^{2k\delta}e^\delta)\delta\nonumber\\
                            &=&\int_0^\infty (\sigma^2-\theta^2 e^{2x})^+\dx+\epsilon, \label{S_int}
\end{eqnarray}
where $\epsilon\rightarrow 0$ as $\alpha\beta\rightarrow\infty$. Observe that by the growth condition 
imposed on $S=S(\alpha\beta)$ and because of $\delta\sim\frac{1}{\alpha\beta}$, it holds 
$\limsup_{\alpha\beta\rightarrow\infty}S\cdot\delta<\infty$ so that transition to a Riemann
integral is allowed; for later reference still note that the water level $\sigma^2=\sigma^2(\alpha\beta)$ 
also remains bounded as $\alpha\beta\rightarrow\infty$. Evaluation of the integral yields
\begin{equation}
  S\doteq\frac{\alpha\beta}{2}\theta^2\cdot\left(\frac{\sigma^2}{\theta^2}
                       \ln\frac{\sigma^2}{\theta^2}-\frac{\sigma^2}{\theta^2}+1\right). \label{S_doteq}
\end{equation}

The maximum in Eq.~(\ref{capacity}) is achieved when the components $X_k$ of the input vector 
$\boldsymbol{X}^K$ are independent $\sim \mathcal{N}(0,\sigma^2-\nu_k^2)$ and the capacity (in nats) 
becomes
\begin{equation}
  C=C(S)=\sum_{k=0}^{K-1}\frac{1}{2}\ln\left(1+\frac{\sigma^2-\nu_k^2}{\nu_k^2}\right). \label{C1}   
\end{equation}
Since $\sigma^2$ eventually remains bounded, transition to a Riemann integral in the next 
equations is allowed and we get
\begin{eqnarray*}
  C\cdot\delta&=&\sum_{k=0}^{K-1}\frac{1}{2}\ln\left(\frac{\sigma^2}{\theta^2}e^{-2k\delta}
                                                                          e^{-\delta}\right)\delta\\
	 &=&\int_0^\infty \left[\frac{1}{2}\ln\left(\frac{\sigma^2}{\theta^2}e^{-2x}\right)\right]^
	                                                                             +\dx+\epsilon\\
	 &=&\frac{1}{2}\left(\frac{1}{2}\ln\frac{\sigma^2}{\theta^2}\right)^2+\epsilon,
\end{eqnarray*}
where $\epsilon\rightarrow 0$ as $\alpha\beta\rightarrow\infty$. Thus it holds 
\begin{equation}
  C\doteq\frac{\alpha\beta}{2}\left(\frac{1}{2}\ln\frac{\sigma^2}{\theta^2}\right)^2.  \label{C1a}
\end{equation}
Eq. (\ref{S_doteq}) is equivalent to
\[
  \frac{\sigma^2}{e\theta^2}\ln\frac{\sigma^2}{e\theta^2}=e^{-1}(s-1)+\epsilon_1,
\]
where $s=S/[(\alpha\beta/2)\theta^2]$ and $\epsilon_1\rightarrow 0$ as $\alpha\beta\rightarrow\infty$. By 
means of the Lambert $W$ function \cite{Corless} (actually its principal branch $W_0$, see 
Fig.~\ref{Figure_6}) which is the uniquely determined analytic function satisfying 
$W(x)\exp[W(x)]=x\:\:\mbox{for all}\:\:x\in[-e^{-1},\infty)$ and $W(0)=0$, we get after a computation
\[
  \frac{1}{2}\left(\frac{1}{2}\ln\frac{\sigma^2}{\theta^2}\right)^2=
                                                \frac{1}{2}\left[w_0(s+\epsilon_1')\right]^2,
\]
where we have put $w_0(x)=\frac{1}{2}[1+W((x-1)/e)],\,x\ge 0$, and $\epsilon_1'=e\epsilon_1$.

\begin{figure}
\centering
\includegraphics[width=3.5in]{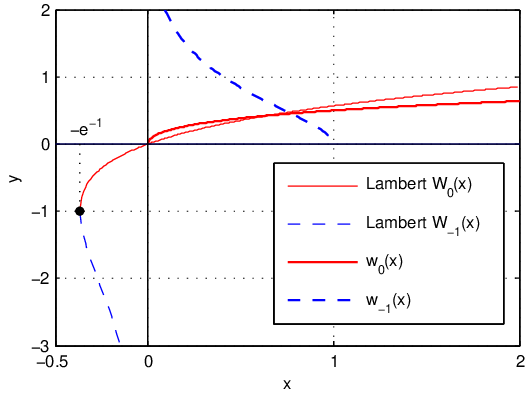}
\caption{Lambert $W$ function: branches $W_0(x),\,W_{-1}(x)$ and related functions $w_0(x),\,w_{-1}(x).$}
\label{Figure_6}
\end{figure}

Because of Eq. (\ref{C1a}), this gives rise to
\[
  \frac{C}{\alpha\beta}=\frac{1}{2}\left[w_0(s+\epsilon_1')\right]^2+\epsilon_2,
\]
where $\epsilon_2\rightarrow 0$ as $\alpha\beta\rightarrow\infty$. Unlike $w_0(x)$, which has a 
vertical tangent at $x=0$, the function $(w_0(x))^2$ has a continuous and bounded derivative in every
closed interval $[0,s_0]\subset[0,\infty)$. Consequently, by the mean value theorem we obtain 
$C/(\alpha\beta)=\frac{1}{2}\left[w_0(s)\right]^2+\epsilon_1''+\epsilon_2$ where 
$\epsilon_1''+\epsilon_2\rightarrow 0$ as $\alpha\beta\rightarrow\infty$. After transition from nats to
bits ($1\:\mbox{nat}= \log_2\!e\:\:\mbox{bit}$), Eq. (\ref{C1b}) is obtained.
\end{proof}

We discuss the case $0<S=S(\alpha\beta)\propto\alpha\beta$ in more detail. First, if $\beta>0$ is held 
constant, then $S=\alpha\beta S_1 =2\pi\alpha P,\,P=\beta S_1/(2\pi)$. Suppose that an individual input 
signal has maximum duration $2\pi\alpha$ (cf.~Fig.~\ref{Figure_9}) and signals are sent every time 
$2\pi\alpha$. Then no ISI occurs and the capacity is approximately $C(S)/(2\pi\alpha)$, where $C(S)$ is 
given by the dotted Eq.~\eqref{C1b}; forming the limit 
$\bar{C}\triangleq\lim_{\alpha\rightarrow\infty}C(S)/(2\pi\alpha)$ turns Eq.~(\ref{C1b}) into the true 
equation
\begin{equation}
   \bar{C}=\frac{1}{2\pi}\frac{\beta}{2}\left[w_0\left(2\pi\frac{P}{(\beta/2)\theta^2}
                                           \right)\right]^2\log_2 e\quad\mbox{(bit/s).} \label{HCh_rate}
\end{equation}
Next, let $\beta\rightarrow\infty$. Since $w_0(0)=0$ and $(w_0(x))^2$ is differentiable at $x=0$ with 
derivative $1/2$, it follows that
\begin{equation}
   \bar{C}\rightarrow\frac{P}{2\theta^2}\log_2 e\:\:\mbox{(bit/s)}.   \label{beta_lim}
\end{equation}

\begin{figure}
\centering
\includegraphics[width=3.5in]{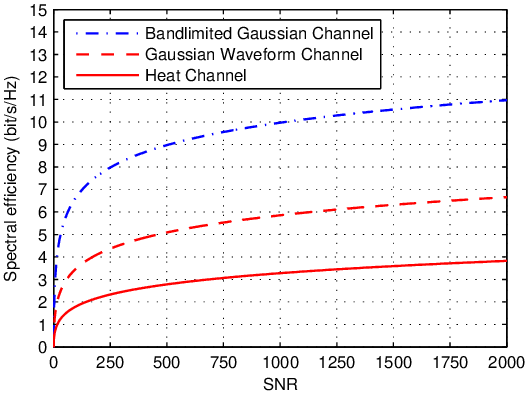}
\caption{Spectral efficiencies of heat channel, bandlimited Gaussian channel, and Gaussian waveform 
channel as a function of SNR.}
\label{Figure_7}
\end{figure}

\begin{figure}
\centering
\includegraphics[width=3.5in]{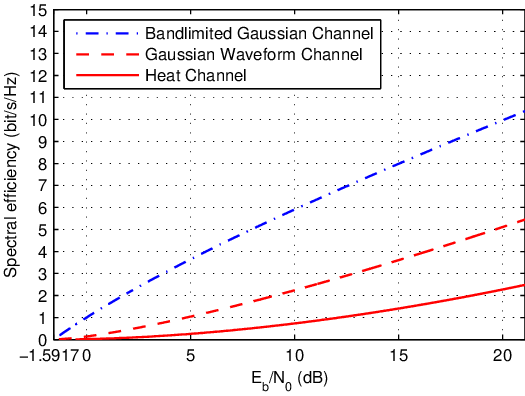}
\caption{Spectral efficiencies of heat channel, bandlimited Gaussian channel, and Gaussian waveform 
channel plotted against $10\log_{10}E_\mathrm{b}/N_0$, where $E_\mathrm{b}$ is average input energy per 
bit and $N_0$ is one-sided noise PSD of the AWGN.}
\label{Figure_8}
\end{figure}

Finally, we compare (\ref{HCh_rate}) with the capacity of the bandlimited Gaussian channel of bandwidth 
$W$ and one-sided noise PSD $N_0$ given by Shannon's classic formula \cite{Shannon1948}
\begin{equation}
  C=W\log_2\left(1+\frac{P}{WN_0}\right)\:\:\mbox{(bit/s)}.    \label{class_Shan}
\end{equation}
Rewrite the latter equation as $C(\mathrm{SNR})=W\log_2(1+\mathrm{SNR})$ where $\mathrm{SNR}=P/(WN_0)$. 
In case of the heat channel, it is consistent to put $\mathrm{SNR}=P/(WN_0)$ where $W=\beta/2$, 
$N_0=2\theta^2$ is the \textit{one}-sided noise PSD, and to rewrite Eq.~\eqref{HCh_rate} as 
$C(\mathrm{SNR})=(2\pi)^{-1}W\,\left[w_0\left(4\pi\,\mathrm{SNR}\right)\right]^2\log_2 e$. In 
Figs.~\ref{Figure_7} and \ref{Figure_8}, the corresponding spectral efficiencies $C(\mathrm{SNR})/W$ are 
plotted as a function of SNR or $E_\mathrm{b}/N_0$, respectively. Although the spectral efficiency of 
the heat channel rapidly falls behind that of the bandlimited Gaussian channel, we observe that the 
capacity limit in (\ref{beta_lim}) is exactly the same as for a Gaussian channel with infinite 
bandwidth, average input power $P$ and one-sided noise PSD $N_0=2\theta^2$; cf.~\cite[Eq.~(9.63)]{Cover}.

\subsection{Optimal Signaling for the Continuous-Time Channel}\label{Section_III-B}
We return to our original model of the heat channel as given in Def.~\ref{DefHC}. For a fixed average 
input energy $S>0$ and noise PSD $\theta^2>0$, the capacity achieving (optimal) input signals are 
now waveforms
\begin{equation}
  f(t)=\sum_{k=0}^{K-1}x_k\,(D_\gamma \psi_k)(t),\,t\in\mathbb{R}, \label{opt_input}
\end{equation}
where the coefficients $x_k$ are realizations of independent Gaussian random variables 
$X_k\sim \mathcal{N}(0,\sigma^2-\nu_k^2),k=0,\ldots,K-1,$ as in the proof of Th.~\ref{form_1}. 
The corresponding (``optimal") filter output signals are
\begin{equation}
  g(t)=\sum_{k=0}^{K-1}y_k\,(D_\gamma \psi_k)(t),\,t\in\mathbb{R}, \label{opt_output}
\end{equation}
where the coefficients $y_k=\rho^{k+\frac{1}{2}}x_k$ are realizations of independent Gaussian random 
variables $Y_k\sim\mathcal{N}(0,\sigma_k^2-\theta^2)$, $\sigma_k^2=\sigma^2\rho^{2k+1},k=0,\ldots,K-1$. 

\begin{figure}
\centering
\includegraphics[width=12cm]{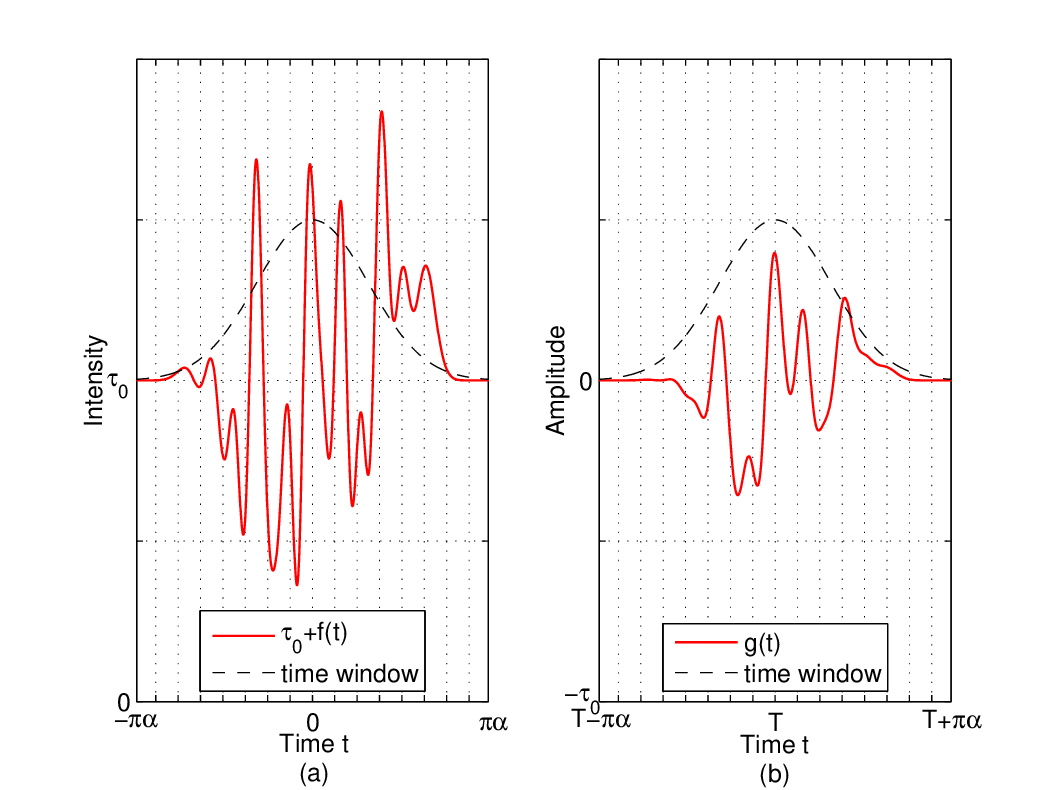}
\caption{Optimal waveforms for the heat channel with parameters 
$\alpha=50\,\mathrm{ps},\,\beta=200\,\mathrm{GHz},\,\mathrm{SNR}=159.1(\approx 1000/2\pi)$.; \emph{time 
window} refers to Gaussian prefactor in \eqref{TFLO}. (a)~Physical input signal $\tau_0+f(t)$ to an 
optical fiber. 
% $\tau=2$ here 
(b)~Corresponding filter output signal $g(t)$ after transmission time $T$.}
\label{Figure_9}
\end{figure}

The following numerical example may serve as illustration. In Fig.~\ref{Figure_9}, a pair of optimal 
filter input/output signals is depicted. Since transmission through an optical fiber by intensity 
modulation is supposed, the input signal $f(t)$ is modified as described in Section~\ref{Section_II-C}. 
By means of Eq.~\eqref{C1b}, the capacity is found to be approximately 64.75~bits per transmission 
(numerical computation gives the exact value of 64.59). Since the effective duration of the fiber input 
signal is less than $2\pi\alpha=0.3141\,\mathrm{ns}$, a sequence of such random pulses sent 
every time $2\pi\alpha$ would transmit approximately 206.1~Gbit/s---in contrast to 50 Gbit/s as inferred
from \eqref{RuleOfThumb}. In the present example, $K=30$ coefficients per waveform are needed.

Now, let $\alpha\beta\rightarrow\infty$. Under the assumption of Th.~\ref{form_1}, the number $K$ of 
coefficients (or active subchannels) is found to be
\begin{equation}
   K\doteq \frac{\alpha\beta}{2}\ln\frac{\sigma^2}{\theta^2}\:. \label{K}
\end{equation}
For the sake of simplicity, assume further that $0<S(\alpha\beta)\propto\alpha\beta$. Then, by 
(\ref{S_int}), we infer that $\sigma^2$ approaches a finite water level $\bar{\sigma}^2>\theta^2$ as 
$\alpha\beta\rightarrow\infty$. As a consequence, $K\rightarrow\infty$ and 
$\sigma_k^2-\theta^2\rightarrow\bar{\sigma}_k^2-\theta^2$ where we have put 
$\bar{\sigma}_k^2=\bar{\sigma}^2\rho^{2k+1}$. When $\bar{\sigma}^2$ is large compared to $\theta^2$, the 
bias $\theta^2$ in $\bar{\sigma}_k^2-\theta^2$ may be neglected so that the optimal filter output signal 
(\ref{opt_output}) comes close to a white Gaussian noise response; cf. (\ref{KL_RD}) (below) and 
Appendix~A. Then, the signal model of Section~\ref{Section_II-D} is almost met and we may take (and 
shall do so in the sequel) the time-frequency product $\alpha\beta$ as DOFs of optimal filter output 
signals.

\subsection{Waterfilling Theorem for the Heat Channel}
By means of a Szeg\H{o} theorem (namely Th.~\ref{TheoremAppendixB} in Appendix~B), the 
above waterfilling solution carries over to the time-frequency plane. The classic waterfilling 
solution for the capacity of a power-constraint additive Gaussian noise channel goes back to 
Shannon \cite{Shannon1949} and has been stated and proved by Gallager \cite{Gallager} in full generality.

Referring to Gallager's capacity theorem \cite[Th.~8.5.1]{Gallager} in the form given in 
\cite[(9.97)]{Cover}, we define
\[
    N(t,\omega)=\frac{\theta^2}{2\pi}\cdot(\cosh\delta)
                  \exp\left(\frac{t^2}{\alpha^2}+\frac{\omega^2}{\beta^2}\right).
\]
Now we are in a position to state
\begin{theorem}\label{form_1a} Under the same assumption on the average input energy $S$ as in 
Th.~\ref{form_1}, the capacity (in bits per transmission) of the heat channel is given by
\begin{equation}
   C\doteq \frac{1}{2\pi}\int\!\!\!\int_{\mathbb{R}^2}\frac{1}{2}
       \log_2\left(1+\frac{(\nu-N(t,\omega))^+}{N(t,\omega)}\right)\dtdomega,
\end{equation}
where $\nu$ is chosen so that
\begin{equation}
   S\doteq\int\!\!\!\int_{\mathbb{R}^2}(\nu-N(t,\omega))^+\dt\domega.
\end{equation}
\end{theorem}
\begin{proof}[Proof:]
See Appendix B. 
\end{proof}

Note that the bivariate function $N(t,\omega)$ is proportional to the reciprocal
WVS $\boldsymbol{\Phi}(t,\omega)$ in (\ref{WVS}). Since $N(t,\omega)$ has the form of a ``cup," 
Th.~\ref{form_1a} is a waterfilling theorem in a very real sense.

When $\alpha\rightarrow\infty$, the time-varying heat channel appears \cite{Ham2013} to tend towards 
an (LTI) waveform channel 
according to Gallager's model\footnote{In the statement of \cite[Th.~8.5.1]{Gallager} the crucial 
assumption ``$T\rightarrow\infty$" is missing. As a consequence, in \cite[Fig.~8.5.1]{Gallager} the 
restriction of the input $x(t)$ to a bounded time interval $(-T/2,T/2)$ would drop out. Therefore 
in Fig.~\ref{Figure_2}(a) any time constraint on the input is omitted.} with LTI filter with Gaussian 
impulse response (\ref{h1}) (we shall call it Gaussian waveform channel). It is therefore interesting to 
compare Th.~\ref{form_1a} with \cite[Th.~8.5.1]{Gallager} when applied to that particular waveform 
channel (with AWGN of noise PSD $N_0/2=\theta^2$). According to Gallager's theorem, the capacity $C$ 
(in bits per second) for input \emph{power} $S$ is given parametrically by (we stick to the notations in 
\cite{Gallager})
\begin{eqnarray*}
   C&=&\int_{-\infty}^\infty\frac{1}{2}
            \left[\log_2\frac{|H_1(f)|^2B}{N_0/2}\right]^+\df\\
   S&=&\int_{-\infty}^\infty\left[
            B-\frac{N_0/2}{|H_1(f)|^2}\right]^+\df,	    
\end{eqnarray*}
where $H_1(f)=\int_{-\infty}^{\infty}e^{-2\pi if t}h_1(t)\dt$ is the frequency response of the filter. 
For the function $h_1(t)$ at hand, we obtain
\begin{eqnarray}
   C&=&\frac{1}{2\pi}\int_{-\infty}^\infty\frac{1}{2}
            \log_2\left(1+\frac{(\nu-N_1(\omega))^+}{N_1(\omega)}\right)\domega\label{C_Gall}\\
   S&=&\int_{-\infty}^\infty(\nu-N_1(\omega))^+\domega,\label{P_Gall}  	    
\end{eqnarray}
where $\nu$ is the parameter, $\omega$ is angular frequency, and
\begin{equation}
   N_1(\omega)=\frac{\theta^2}{2\pi}\cdot \exp\left(\frac{\omega^2}{\beta^2}\right). \label{N1}
\end{equation}
We observe perfect formal analogy between the waterfilling formulas (\ref{C_Gall}), (\ref{P_Gall}) 
and those in Th.~\ref{form_1a}. Moreover, $N(t,\omega)$ tends to $N_1(\omega)$ as 
$\alpha\rightarrow\infty$ for any $t,\omega$ held constant. In Figs.~\ref{Figure_7} and \ref{Figure_8}, 
the spectral efficiency $C/W$ of the Gaussian waveform channel is plotted as a function of SNR or 
$E_\mathrm{b}/N_0$, respectively.

\section{Rate Distortion Function for a Related Nonstationary Source}\label{Section_IV}
For a well-rounded treatment of the capacity problem for the heat channel it is expedient to investigate 
a dual problem, which is a topic of rate distortion theory. 
To this end, consider the nonstationary source given by the nonstationary zero-mean Gaussian process 
defined by the Karhunen-Lo\`{e}ve expansion
\begin{equation}
  X(t)=\sum_{k=0}^\infty X_k\,(D_\gamma \psi_k)(t),\,t\in\mathbb{R},\label{KL_RD}
\end{equation}
where the coefficients $X_k,\,k=0,1,\ldots,$ are independent Gaussian random variables 
$\sim\mathcal{N}(0,\sigma_k^2)$ of variance $\sigma_k^2=\sigma^2\rho^{2k+1},\,\sigma>0$. It is the 
response of filter $\boldsymbol{P}_\delta^{(\gamma)}$ on white Gaussian noise; cf. 
(\ref{noise_resp}) in Appendix~A. In Fig.~\ref{Figure_4}, the area beneath the curve 
$y=\sigma^2e^{-2x}$ corresponds to the average energy
\begin{equation}
  E=\sum_{k=0}^\infty\sigma_k^2=\frac{\sigma^2\rho}{1-\rho^2}
                                     \doteq\frac{\alpha\beta}{2}\sigma^2\label{av_en_RD}
\end{equation}
of the Gaussian process (\ref{KL_RD}). The parameter $\theta^2$ in Fig.~\ref{Figure_4} will now have 
the interpretation of a ``(ground-)water table." In this section, information will be measured in nats.
\subsection{Rate Distortion Function in Closed Form}
Substitute the continuous-time Gaussian process $\{X(t),\,t\in\mathbb{R}\}$ in (\ref{KL_RD}) by the
sequence of coefficient random variables $\boldsymbol{X}=X_0,X_1,\ldots\,$. For an estimate 
$\boldsymbol{\hat{X}}=\hat{X}_0,\hat{X}_1,\ldots$ of $\boldsymbol{X}$ we take the mean-square error 
$D=\mathbb{E}\{\sum_{k=0}^\infty(X_k-\hat{X}_k)^2\}$ as distortion measure.

The function $y=w_{-1}(x),\,0<x\le1,$ occurring in the next theorem is the inverse function of 
$y=(2x+1)e^{-2x},\,x\ge 0$ (see also Fig.~\ref{Figure_6}). The Landau symbol $\Omega(\cdot)$ is 
defined for any two functions as in Def.~\ref{def_2} as follows: $A(x)=\Omega(B(x))$ as 
$x\rightarrow\infty$ if $B(x)>0$ and $\liminf_{x\rightarrow\infty}A(x)/B(x)>0$.
\begin{theorem}\label{R(D)} Assume that the foregoing average distortion $D$ depends on 
$\alpha\beta$ such that $D(\alpha\beta)=\Omega(\alpha\beta)$ as $\alpha\beta\rightarrow\infty$. Then the 
rate distortion function for the nonstationary source (\ref{KL_RD}) satisfies 
\begin{equation}R(D)\doteq\frac{\alpha\beta}{2}\left[w_{-1}\left(\frac{D}{(\alpha\beta/2)\sigma^2}
                                                                       \right)\right]^2   \label{R_D}
\end{equation}
if $D\le (\alpha\beta/2)\sigma^2$, and $R(D)\doteq 0$ otherwise. The rate is measured in nats per 
realization of the source.
\end{theorem}
\begin{proof}[Proof:] Let $E$ be the average energy (\ref{av_en_RD}) of the Gaussian process (\ref{KL_RD}).

First, assume $D\le E$. The reverse waterfilling argument for a finite number of 
independent Gaussian sources \cite{Cover} carries over to our case without changes resulting in a finite 
collection of Gaussian sources $X_0,\ldots,X_{K-1}$ where 
$K=\max\{k\in\mathbb{N};\sigma_{k-1}^2>\theta^2\}$ and the water level $\theta^2>0$ is defined by 
the condition
\begin{equation}
  D=\sum_{k=0}^\infty \min\{\theta^2,\sigma_k^2\},  \label{def_theta}
\end{equation}
(cf. Fig.~\ref{Figure_4}, where $D=E_\mathrm{dist1}+E_\mathrm{dist2}$). Consequently, 
\begin{eqnarray}
  D\cdot\delta&=&\sum_{k=0}^\infty \min\{\theta^2,\sigma^2e^{-2k\delta}e^{-\delta}\}\delta\nonumber\\
         &=&\int_0^\infty \min\{\theta^2,\sigma^2e^{-2x}\}\dx+\epsilon,  \label{D_int}
\end{eqnarray}
where $\epsilon\rightarrow0$ as $\alpha\beta\rightarrow\infty$. Observe that by the growth condition 
imposed on  $D$ and since $\delta\sim\frac{1}{\alpha\beta}$, the water level $\theta^2$ eventually 
remains above a positive lower bound as $\alpha\beta\rightarrow\infty$. Evaluation of the integral 
yields
\begin{equation}
    D\doteq\frac{\alpha\beta}{2}\sigma^2\cdot\left(\frac{\theta^2}{\sigma^2}
                     -\frac{\theta^2}{\sigma^2}\ln\frac{\theta^2}{\sigma^2}\right).  \label{D_doteq}
\end{equation}

The rate distortion function is parametrically given by \cite{Cover}
\begin{equation}
  R = \sum_{k=0}^{K-1}\frac{1}{2}\ln\frac{\sigma_k^2}{\theta^2}. \label{R1}
\end{equation}
The RHSs of Eqs. (\ref{R1}) and (\ref{C1}) agree. Since $\frac{1}{\theta^2}$ is eventually 
finitely upper bounded, transition to a Riemann integral is allowed and we obtain exactly as in the 
proof of (\ref{C1a}) that
\begin{equation}
  R\doteq\frac{\alpha\beta}{2}\left(\frac{1}{2}\ln\frac{\sigma^2}{\theta^2}\right)^2.  \label{R1a}   
\end{equation}

Eq. (\ref{D_doteq}) is equivalent to  
\[
  \frac{\theta^2}{e\sigma^2}\ln\frac{\theta^2}{e\sigma^2}=-e^{-1}d+\epsilon_1,
\]
where $d=D/[(\alpha\beta/2)\sigma^2]$ and $\epsilon_1\rightarrow 0$ as $\alpha\beta\rightarrow\infty$. 
By means of the branch $W_{-1}$ of the Lambert $W$ function \cite{Corless}, which is the uniquely 
determined analytic function satisfying 
$W_{-1}(x)\exp[W_{-1}(x)]=x\:\:\mbox{for all}\:\: x\in[-e^{-1},0)$ and 
$W_{-1}(-e^{-1})=-1,\,W_{-1}(x)\rightarrow -\infty$ as $x\rightarrow 0-$ (see Fig.~\ref{Figure_6}), we 
get after a computation
\[
  \frac{1}{2}\left(\frac{1}{2}\ln\frac{\sigma^2}{\theta^2}\right)^2=
                                                      \frac{1}{2}\left[w_{-1}(d+\epsilon_1')\right]^2,
\]
where we have put $w_{-1}(x)=\frac{1}{2}[-1-W_{-1}(-x/e)],\,0<x\le 1$, and $\epsilon_1'=-e\epsilon_1$. 
Because of Eq. (\ref{R1a}), this gives rise to
\[
  \frac{R}{\alpha\beta}=\frac{1}{2}\left[w_{-1}(d+\epsilon_1')\right]^2+\epsilon_2,
\]
where $\epsilon_2\rightarrow 0$ as $\alpha\beta\rightarrow\infty$. Unlike $w_{-1}(x)$, which has a
vertical tangent at $x=1$,  the function $(w_{-1}(x))^2$ has a continuous and bounded derivative in any 
closed interval $[d_0,1]\subset(0,1]$. Consequently, by the mean value theorem 
$R/(\alpha\beta)=\frac{1}{2}\left[w_{-1}(d)\right]^2+\epsilon_1''+\epsilon_2$ where 
$\epsilon_1''+\epsilon_2\rightarrow 0$ as $\alpha\beta\rightarrow\infty$ uniformly for all $d\in[d_0,1]$.
This proves the first part of the theorem.

Now, suppose that $D>E$. Since $\mathbb{E}\{\sum_{k=0}^\infty(X_k-0)^2\}=E<D$, the constant sequence 
$\boldsymbol{\hat{X}}=0,0,\ldots$ is a sufficient estimate. Since there is no uncertainty about 
the members of that deterministic sequence, no information needs to be supplied; thus, $R=0$. This 
proves the second part of the theorem.
\end{proof}

\subsection{Reverse Waterfilling in the Time-Frequency Plane}
Before continuing with our main theme, we present a parametric representation of the rate distortion 
function occurring in Th.~\ref{R(D)} since the means for its proof---Th.~\ref{TheoremAppendixB} in 
Appendix~B---are now available. This representation may be viewed as an extension to the time-frequency 
plane of the classic method of reverse waterfilling (cf., e.g., \cite{Berger}, \cite{Cover}) due to 
Kolmogorov \cite{Kolmogorov}. It turns out that the part of the PSD in \cite[Th.~4.5.4]{Berger} is now 
taken by the WVS $\boldsymbol{\Phi}(t,\omega)$ in (\ref{WVS}). We obtain
\begin{theorem}\label{R(D)_param} The rate distortion function $R(D)$ for the nonstationary 
source (\ref{KL_RD}) has in the interval $0<D\le (\alpha\beta/2)\sigma^2$ the parametric representation
\begin{eqnarray*}
   R_\lambda&\doteq&\frac{1}{2\pi}\int\!\!\!\int_{\mathbb{R}^2}
       \max\left\{0,\frac{1}{2}\ln\frac{\boldsymbol{\Phi}(t,\omega)}{\lambda}\right\}\dt\domega\\
   D_\lambda&\doteq&\int\!\!\!\int_{\mathbb{R}^2}
            \min\left\{\lambda,\boldsymbol{\Phi}(t,\omega)\right\}\dtdomega.
\end{eqnarray*}
\end{theorem} 
\begin{proof}[Proof:]
See Appendix B.
\end{proof}

Note that
\[
  \int\!\!\!\int_{\mathbb{R}^2}\boldsymbol{\Phi}(t,\omega)\dt\domega
           =\frac{\alpha\beta}{2}\frac{\sigma^2}{\cosh\delta}\doteq\frac{\alpha\beta}{2}\sigma^2
\]
is the average energy (\ref{av_en_RD}) of the nonstationary Gaussian process (\ref{KL_RD}) (as it 
should be). We observe that the representation in Th.~\ref{R(D)_param} is in perfect analogy to the 
parametric representation \cite[Eqs.~(4.5.51),~(4.5.52)]{Berger} of the $R(D)$ function for a 
continuous-time stationary Gaussian source.

\section{Channel Capacity in Terms of Channel Output, and Relation to Detection Theory}\label{Section_V}
Now, we adopt the perspective of the receiver. This will result in a second closed-form capacity formula 
for the heat channel, now in terms of elementary functions and akin to the classic Shannon formula 
(\ref{class_Shan}) for the capacity of a bandlimited Gaussian channel. Moreover, we shall find a 
parallel to the well-known I-MMSE relationship \cite{GSV}. Implications 
for the capacity of multiple-input multiple-output (MIMO) systems will be indicated. In the present 
section, it is convenient to use natural logarithms; therefore, information is measured in nats.

\subsection{Channel Capacity---Second Formula in Closed Form}\label{Section_V-A}
For any fixed average input energy $S$ the capacity achieving input signals to the 
continuous-time heat channel are waveforms $f(t)=\sum_{k=0}^{K-1}x_k\,(D_\gamma \psi_k)(t)$, where 
the coefficients $x_k$ are realizations of independent Gaussian random variables 
$X_k\sim\mathcal{N}(0,\sigma^2-\nu_k^2)$ (where $K$, $\sigma$ and $\nu_k$ are same as in the proof of 
Th.~\ref{form_1}). In the next theorem, the capacity of the heat channel will be expressed in terms 
of the average energy $\hat{E}_\mathrm{out}$ of the detected perturbed filter output signal 
$\hat{g}(t)=\sum_{k=0}^{K-1}\hat{y}_k\,(D_\gamma \psi_k)(t)$, or rather its coefficients 
$\hat{y}_k=\rho^{k+\frac{1}{2}}x_k+n_k$ obtained by optimal detection (through matched filters). 
Then the detection errors $n_k$ are realizations of $\mathrm{i.i.d.}$ random variables 
$N_k\sim\mathcal{N}(0,\theta^2)$ as in Section \ref{Section_II}.
 
\begin{theorem}\label{form_2} Assume that the average energy $\hat{E}_{\mathrm{out}}$ of the 
detected perturbed filter output signal depends on $\alpha\beta$ such that 
$\hat{E}_{\mathrm{out}}(\alpha\beta)=O(\alpha\beta)$ as $\alpha\beta\rightarrow\infty$. Then the 
capacity (in nats per transmission) of the heat channel is given by
\begin{equation}C\doteq\frac{\alpha\beta}{2}
       \left[\ln\sqrt{1+\frac{\hat{E}_{\mathrm{out}}}{(\alpha\beta/2)\theta^2}}\,\right]^2.\label{C2}
\end{equation}
\end{theorem}
\begin{proof}[Proof:]
Rewrite Eq. (\ref{C1a}) as
\begin{eqnarray}
  C\doteq\frac{\alpha\beta}{2}\left[\ln\sqrt{1+\frac{(\alpha\beta/2)(\sigma^2-\theta^2)}
	                        {(\alpha\beta/2)\theta^2}}\,\right]^2                \label{C2_right}
\end{eqnarray}
and put $\sigma_k^2=\sigma^2\rho^{2k+1},\,k=0,1,\ldots\,$. In case of capacy achieving (optimal) 
signaling, the average energy of the detected perturbed filter output signal is 
$\hat{E}_\mathrm{out}=E_\mathrm{out}+E_\mathrm{err}$ (cf. Fig.~\ref{Figure_4}) where 
$E_\mathrm{out}=\sum_{k=0}^{K-1}(\sigma_k^2-\theta^2)$ is the average energy of the filter output 
signal, $E_\mathrm{err}=\sum_{k=0}^{K-1}\theta^2$ is the average energy of the (total) detection error, 
and $K$ is given by $K=\max\{k\in\mathbb{N};\sigma_{k-1}^2>\theta^2\}$ (coinciding with the number $K$ 
of active subchannels in the proof of Th.~\ref{form_1}). Since 
$\hat{E}_\mathrm{out}=\sum_{k=0}^{K-1}\sigma_k^2$ and $\delta\sim\frac{1}{\alpha\beta}\,
                                                               (\alpha\beta\rightarrow\infty)$, we get
\[
  \hat{E}_\mathrm{out}\cdot\delta = e^{-\delta}\sum_{k=0}^{K-1}\sigma^2e^{-2k\delta}\delta
     =\int_0^{\frac{1}{2}\ln\frac{\sigma^2}{\theta^2}}\sigma^2e^{-2x}\dx+\epsilon
     =\frac{1}{2}(\sigma^2-\theta^2)+\epsilon,
\]
where $\epsilon\rightarrow 0$ as $\alpha\beta\rightarrow\infty$. Observe that by the growth condition 
imposed on $\hat{E}_{\mathrm{out}}$, $\sigma^2$ remains bounded as $\alpha\beta\rightarrow\infty$ 
(justifying in hindsight the transition to a Riemann integral). Hence, 
\begin{equation}
   \hat{E}_\mathrm{out}\doteq\frac{\alpha\beta}{2}(\sigma^2-\theta^2).         \label{ref_sigma}
\end{equation}
Now Eq. (\ref{C2}) follows from Eqs.~(\ref{C2_right}),~(\ref{ref_sigma}).
\end{proof}
Note that for the determination of channel capacity by formula (\ref{C2}), the receiver does 
not need to know the number $K$ of active subchannels beforehand, since, at least in principle, it 
could easily be estimated as accurately as desired from successive optimal channel uses at constant 
average input energy.

In the rest of this section we shall always suppose that $K$ is given by Eq.~(\ref{K}), tacitly assuming 
that the assumptions of Th.~\ref{form_1} (or Th.~\ref{form_2}) are fulfilled.

\subsection{A C-NODE Relationship} \label{Section_V-B}
The setting of the previous subsection gives rise to a vector Gaussian channel
\begin{equation}
  \boldsymbol{Y}^K=\boldsymbol{H}^K\boldsymbol{X}^K+\boldsymbol{N}^K,    \label{VGC2}
\end{equation}
where the matrix $\boldsymbol{H}^K$ is the $K\times K$ diagonal matrix with entries 
$h_{kk}=\rho^{k+\frac{1}{2}},\,k=0,\ldots,K-1$, $\boldsymbol{X}^K$ is a random
input vector satisfying $\mathbb{E}\|\boldsymbol{X}^K\|<\infty$, and the noise vector $\boldsymbol{N}^K$ 
has independent random components $N_k\sim\mathcal{N}(0,\theta^2)$. Recall that the noise is caused by 
errors coming from optimal detection. The noise has average energy $\mathbb{E}\|\boldsymbol{N}^K\|^2$; 
we define the \emph{normalized optimal detection error} (NODE) as
\begin{equation}
  \node(S)=\frac{\mathbb{E}\|\boldsymbol{N}^K\|^2}{\sigma^2}=\frac{K\theta^2}{\sigma^2} \label{nodeDef}
\end{equation}
where $S$ is the average energy of channel input signals (or vectors) and $\sigma^2$ is determined by 
Eq.~\eqref{def_sigma}. Because of Eq.~\eqref{K}, it holds that
\begin{equation}
  \node(S)\doteq\frac{\alpha\beta}{2}\frac{\theta^2}{\sigma^2}\ln\frac{\sigma^2}{\theta^2} \label{nodeS}
\end{equation}
The quantity $\node(S)$ is called \emph{normalized} because 
$\mathbb{E}\|\boldsymbol{N}^K\|^2=K\theta^2\doteq\frac{\alpha\beta}{2}\theta^2\ln\frac{\sigma^2}{\theta^2}$ 
turns into the RHS of Eq.~\eqref{nodeS} after rescaling $\sigma,\theta\leftarrow 1,\theta/\sigma$. 
Notice that the NODE 
as defined in \eqref{nodeDef} is physically dimensionless.

\subsubsection{Motivation and Derivation} \label{Sec_MotivationEtc}
The central result of \cite{GSV} is an identity connecting mutual information with the minimum 
mean-square error (MMSE) of estimation theory (I-MMSE relationship); it reads
\begin{equation}
  \dnachdsnr I(\boldsymbol{X};\sqrt{\snr}\boldsymbol{H}\boldsymbol{X}+\boldsymbol{N})=\frac{1}{2}
                                           \mmse(\snr),                                   \label{I-MMSE}
\end{equation}
where $\boldsymbol{N}$ is a noise vector with independent standard Gaussian components, independent of the random 
vector $\boldsymbol{X}$, $\mathbb{E}\|\boldsymbol{X}\|^2<\infty$, and $\boldsymbol{H}$ is a deterministic matrix of 
appropriate dimension. It is interesting to compare Eq.~\eqref{I-MMSE} with the capacity calculations in our paper. 
We shall denote the inverse matrix of $\boldsymbol{H}^K$ by $\boldsymbol{H}^{-K}$. Since mutual information is 
invariant with respect to invertible linear transformations, we infer for the mutual information 
$I(\boldsymbol{X}^K;\boldsymbol{Y}^K)=I(\boldsymbol{X}^K;\boldsymbol{X}^K+\boldsymbol{H}^{-K}\boldsymbol{N}^K)$ 
occurring in Eq.~\eqref{capacity} that
\begin{equation}
I(\boldsymbol{X}^K;\boldsymbol{Y}^K)   = I\left(\boldsymbol{X}'^K;\frac{\sigma}{\theta}
                                      \boldsymbol{H}^K\boldsymbol{X}'^K+\boldsymbol{N}'^K\right). \label{I_I}
\end{equation}
where the noise vector $\boldsymbol{N}'^K=\theta^{-1}\boldsymbol{N}^K$ has independent standard 
Gaussian components, independent of the random vector $\boldsymbol{X}'^K=\sigma^{-1}\boldsymbol{X}^K$. If we take 
in \eqref{I_I} for $\boldsymbol{X}'^K=(X'_0,\ldots,X'_{K-1})^T$ a random vector with independent components 
$X_k'\sim\mathcal{N}(0,1-(\theta^2/\sigma^2)\rho^{-2k-1})$, then the capacity $C(S)$ of the heat channel 
is achieved  (cf. proof of Th.~\ref{form_1}). Since $C(S)$ depends only on the signal-to-noise ratio 
$\snr=\sigma^2/\theta^2\in[1,\infty)$,\footnote{Since only the portion $\sigma^2-\theta^2$ contributes to 
the signal, $\sigma^2/\theta^2$ is rather a \textit{signal plus noise}-to-noise ratio; we stick to the notation 
``snr" to conform with \cite{GSV}.} we may write (with slight abuse of notation)
\begin{equation}
  C(\snr) = I(\boldsymbol{X}'^K;\sqrt{\snr}\boldsymbol{H}^K
                                     \boldsymbol{X}'^K+\boldsymbol{N}'^K),\label{C_snr}
\end{equation}
which is reminiscent of the mutual information in (\ref{I-MMSE}).

\begin{figure}
\centering
\includegraphics[width=3.5in]{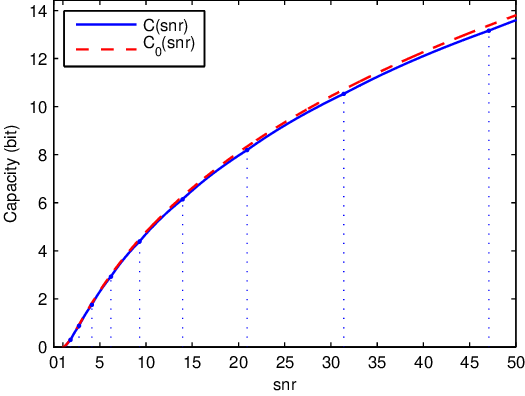}
\caption{Capacity $C(\snr)$ of heat channel, and its approximation $C_0(\snr)$ in case 
$\alpha\beta=5$ (for larger values of $\alpha\beta$, the two curves quickly become indistinguishable in 
the given range of snr). Dotted lines depict snrs where differentiability of $C(\snr)$ breaks 
down.}
\label{Figure_10}
\end{figure}

Now, several problems arise when trying to take the derivate with respect to snr: 1) the probability 
distribution of the input vector $\boldsymbol{X}'^K$ depends on snr (a situation not covered by 
\cite[Th.~2]{GSV}), 2) the function $C(\snr)$ is not differentiable at snrs where a new subchannel is 
added (cf. Fig.~\ref{Figure_10}). To overcome both difficulties, we substitute $C(\snr)$ by its smooth 
approximation
\[
  C_0(\snr)\triangleq\frac{\alpha\beta}{2}\left(\ln\sqrt{\snr}\right)^2
\]
as given by the RHS of Eq.~\eqref{C1a}. Since in Eq.~\eqref{nodeDef}, 
$K=\max\{k\in\mathbb{N};\rho^{-2k+1}<\sigma^2/\theta^2\}$, $\node(S)$ actually only depends on $\snr$ 
and we shall write $\node(\snr)$ instead. Now, we obtain
\begin{theorem}[C-NODE Relationship]\label{C-NODE} For any $\snr\ge1$ it holds that 
\begin{equation}
  \dnachdsnr C_0(\snr)\doteq\frac{1}{2}\node(\snr), \label{C0_node}
\end{equation}
where
\begin{equation}
   \node(\snr)\doteq\frac{\alpha\beta}{2}\frac{\ln\snr}{\snr}.  \label{nodesnr}
\end{equation}
\end{theorem}
\begin{proof}[Proof:]
Eq.~\eqref{nodesnr} is just Eq.~\eqref{nodeS} put in other terms. Taking the derivative of $C_0(\snr)$ 
yields
\[
  \dnachdsnr C_0(\snr)=\frac{\alpha\beta}{2}\frac{1}{2}\frac{\ln\snr}{\snr}\doteq\frac{1}{2}\node(\snr),
\]
thus proving Eq.~\eqref{C0_node}.
\end{proof}

Observe the striking similarity between Eqs. \eqref{C0_node} and \eqref{I-MMSE}. Eq.~\eqref{C0_node} 
establishes a connection between (increase of) capacity and the NODE in the vector Gaussian channel \eqref{VGC2}, so the 
name of theorem. Notice that Th.~\ref{C-NODE} links information theory with \emph{detection} theory just 
as does the I-MMSE relationship \eqref{I-MMSE} with the former and \emph{estimation} theory.

\subsubsection{Discussion}
To recognize the difference between Eqs.~\eqref{I-MMSE} and (\ref{C0_node}), we calculate the MMSE. We
continue to suppose that $\snr=\sigma^2/\theta^2\ge 1$; the transpose of matrix 
$\boldsymbol{H}^K$ will be denoted by $\boldsymbol{H}^{KT}.$ Following \cite{GSV}, given  
\begin{equation}
  \boldsymbol{Y}^K=\sqrt{\snr}\boldsymbol{H}^K\boldsymbol{X}'^K+\boldsymbol{N}'^K, \label{MIMO_GSV}
\end{equation}
the MMSE in estimating $\boldsymbol{H}^K\boldsymbol{X}'^K$is
\begin{eqnarray*}
  \mmse(\snr)
    &=&\mathbb{E}\left\|\boldsymbol{H}^K\boldsymbol{X}'^K-\boldsymbol{H}^K
                                 \widehat{\boldsymbol{X}'^K}\right\|^2\\
    &=&\mathrm{tr}[\boldsymbol{H}^K(\boldsymbol{\Sigma}^{-K}+\snr
                                {\boldsymbol{H}^{KT}}\boldsymbol{H}^K)^{-1}{\boldsymbol{H}^{KT}}],
\end{eqnarray*}
where $\widehat{\boldsymbol{X}'^K}$ is the minimum mean-square estimate of $\boldsymbol{X}'^K$, and 
$\boldsymbol{\Sigma}^{-K}$ is the inverse of the covariance matrix $\boldsymbol{\Sigma}^K$ of 
$\boldsymbol{X}'^K$. If $\boldsymbol{X}'^K$ has independent Gaussian components 
$X_k'\sim\mathcal{N}(0,1-\snr^{-1}\rho^{-2k-1})$ as in Section~\ref{Sec_MotivationEtc}, then 
$\boldsymbol{\Sigma}^K$ is a $K\times K$ diagonal matrix with entries 
$\sigma_{kk}=1-\snr^{-1}\rho^{-2k-1},\,k=0,\ldots,K-1$. A computation yields
\[
  \mmse(\snr)=\sum_{k=0}^{K-1}\snr^{-1}
                                                \left(1-\snr^{-1}\rho^{-2k-1}\right).
\]
When $\alpha\beta$ becomes large, we obtain by transition to a Riemann integral
\begin{eqnarray*}
  \mmse(\snr)\cdot\delta
        &=&\sum_{k=0}^{K-1}\snr^{-1}\left(1-\snr^{-1}\rho^{-2k-1}\right)\delta\\
        &=&\snr^{-1}\int_0^\infty\left(1-\snr^{-1}e^{2x}\right)^+\dx+\epsilon\\
	&=&\frac{1}{2}\frac{\ln\snr}{\snr}
	       -\frac{1}{2}\frac{1}{\snr}\left(1-\frac{1}{\snr}\right)+\epsilon,
\end{eqnarray*}
where $\epsilon\rightarrow 0$ as $\alpha\beta\rightarrow\infty$. Thus,
\[
  \mmse(\snr)\doteq\frac{\alpha\beta}{2}\frac{\ln\snr}{\snr}
	              -\frac{\alpha\beta}{2}\frac{1}{\snr}\left(1-\frac{1}{\snr}\right),
\]
or, using Eq.~(\ref{nodesnr}),
\begin{equation}
  \mmse(\snr)\doteq \node(\snr)
                     -\frac{\alpha\beta}{2}\frac{1}{\snr}\left(1-\frac{1}{\snr}\right).
		                                                                           \label{mmse}
\end{equation}
Finally, averaging with respect to the DOFs $\alpha\beta$ turns the (dotted) equations 
(\ref{nodesnr}),~(\ref{mmse}) into true equations and we get for NODE and MMSE, resp.,
\begin{align*} 
   \overline{\node}(\snr)
           &\triangleq\lim_{\alpha\beta\rightarrow\infty}\frac{\node(\snr)}{\alpha\beta}
	    =\frac{1}{2}\frac{\ln\snr}{\snr},\\
   \overline{\mmse}(\snr)
           &\triangleq\lim_{\alpha\beta\rightarrow\infty}\frac{\mmse(\snr)}{\alpha\beta}
	    =\overline{\node}(\snr)-\frac{1}{2}\frac{1}{\snr}\left(1-\frac{1}{\snr}\right).
\end{align*}

\begin{figure}
\centering
\includegraphics[width=3.5in]{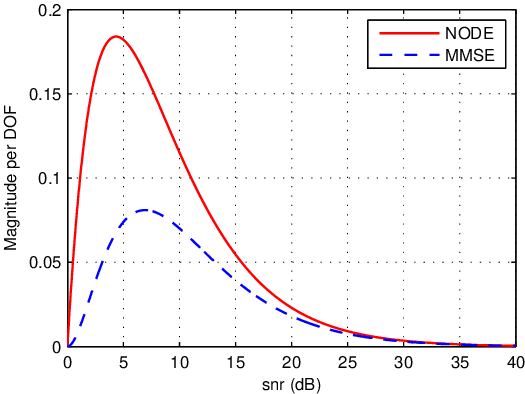}
\caption{Magnitude per DOF of NODE and MMSE as DOFs $\alpha\beta\rightarrow\infty$.}
\label{Figure_11}
\end{figure}

In Fig.~\ref{Figure_11}, $\overline{\node}(\snr)$ and $\overline{\mmse}(\snr)$ are plotted against 
$10\log_{10}\snr$ for $\snr\ge1$. Apparently, the increase of capacity of the vector Gaussian 
channel~\eqref{MIMO_GSV} with growing snr as predicted by the C-NODE relationship of Th.~\ref{C-NODE} 
is significantly higher than anticipated by the I-MMSE relationship~\eqref{I-MMSE}, 
at least in the lower snr region (and at the expense of higher dimension $K$ since $K\rightarrow\infty$ 
as $\alpha\beta\rightarrow\infty$). This observation might also be useful for the 
assessment of the capacity of (high dimensional) MIMO systems, where, by the way, the case of 
snr-dependent input signals has rarely been treated so far (cf., e.g., \cite{GSV}, \cite{Telatar}, 
\cite{LTV}, \cite{PalomarVerdu07}).

\section*{Appendix}
\begin{appendix}

\section{Wigner-Ville Spectrum of Filter Response on White Gaussian Noise}
We model white Gaussian noise of two-sided noise PSD $\sigma^2\in(0,\infty)$ by a sequence of 
stochastic processes $\{\boldsymbol{U}^K(t),t\in\mathbb{R}\},\,K=1,2,\ldots,$ given by their respective 
Karhunen-Lo\`{e}ve expansion
\[
  \boldsymbol{U}^K(t)=\sum_{k=0}^{K-1}U_k\,(D_\gamma \psi_k)(t),\,t\in\mathbb{R},
\]
where $U_0,\ldots,U_{K-1}$ are $\mathrm{i.i.d.}$ Gaussian random variables $\sim\mathcal{N}(0,\sigma^2)$. 
For any $K=1,2,\ldots,$ let the process $\{\boldsymbol{U}^K(t),t\in\mathbb{R}\}$ be the input to filter 
$\boldsymbol{P}_\delta^{(\gamma)}$. By means of representation (\ref{EXP}) it is seen that the 
corresponding filter output tends as $K\rightarrow\infty$ to the stochastic process
\begin{equation}
  X(t)=\sum_{k=0}^\infty\rho^{k+\frac{1}{2}}U_k\,(D_\gamma \psi_k)(t),\,t\in\mathbb{R}.\label{noise_resp}
\end{equation}
We interprete $\{X(t),t\in\mathbb{R}\}$ as filter response on white Gaussian noise.

Since any realization $x(t)$ of $\{X(t)\}$ is almost surely in $L^2(\mathbb{R})$, the Wigner distribution 
\cite{Groch}
\[ 
  (Wx)(t,\omega)=\frac{1}{2\pi}\int_{-\infty}^\infty e^{-i\omega t'}
        x\left(t+\frac{t'}{2}\right)\overline{x\left(t-\frac{t'}{2}\right)}\dt'
\]
may be computed. By taking the ensemble average, we obtain the WVS \cite{Flandrin} of process $\{X(t)\}$,
\begin{eqnarray}
  \boldsymbol{\Phi}(t,\omega)&=&\mathbb{E}[(WX)(t,\omega)] \nonumber\\
    &=&\frac{1}{2\pi}\int_{-\infty}^{\infty}e^{-i\omega t'}r\left(t+\frac{t'}{2},t-\frac{t'}{2}
                                                                             \right)\dt',\label{def_WVS}
\end{eqnarray}
where $r(t_1,t_2)=\mathbb{E}[X(t_1)\overline{X(t_2)}]$ is the autocorrelation function. 
The kernel of operator $\boldsymbol{P}_\delta^{(\gamma)}$ has for arbitrary parameters 
$\gamma,\delta>0$ two alternative representations (following from a generalization of Mehler's formula \cite{Ham2004}),
\begin{align*}
\lefteqn{P_\delta^{(\gamma)}(x,y)}\\
    &=\sum_{k=0}^\infty\rho^{k+\frac{1}{2}}(D_\gamma \psi_k)(x)(D_\gamma \psi_k)(y)\\
    &=\frac{1}{\gamma\sqrt{2\pi\sinh\delta}}
         \exp\left\{-\frac{1}{4\gamma^2}\left[\coth\left(\frac{\delta}{2}\right)(x-y)^2
                                       +\tanh\left(\frac{\delta}{2}\right)(x+y)^2\right]\right\}.
\end{align*}
We infer by the first representation that $r(t_1,t_2)=\sigma^2 P_{2\delta}^{(\gamma)}(t_1,t_2)$. Then, by 
means of the second representation, the integral in (\ref{def_WVS}) is readily evaluated; we obtain
\[
  \boldsymbol{\Phi}(t,\omega)=\frac{\sigma^2}{2\pi}\cdot\frac{1}{\cosh\delta}
                      \exp\left(-\frac{t^2}{\alpha^2}-\frac{\omega^2}{\beta^2}\right).
\]

\section{Proofs of Theorems~\ref{form_1a} and \ref{R(D)_param}}
For a linear operator $A:L^2(\mathbb{R})\rightarrow L^2(\mathbb{R})$ the Weyl symbol $\sigma_A(x,\xi)$---when 
existing \cite{Groch}---is defined by \cite{KoHlaw}, \cite{JZ} 
\begin{eqnarray}
(Af)(x)=\frac{1}{2\pi}\int\!\!\!\int_{\mathbb{R}^2}\sigma_A
                         \left(\frac{x+y}{2},\xi\right)e^{i(x-y)\xi}f(y)\dy\,d\xi.\label{kernel_2}
\end{eqnarray}
The linear map $A\mapsto\sigma_A(x,\xi)$ (or rather its inverse) is called Weyl correspondence. For example, 
the operator $A=\boldsymbol{P}_{2\delta}^{(\gamma)}$ has the Weyl symbol \cite{Ham2004}
\begin{eqnarray}
   \sigma_A(x,\xi)&=&\frac{1}{\cosh\delta}
                             e^{-(\tanh\delta)(\gamma^{-2}x^2+\gamma^2\xi^2)}    \label{Weyl_symbol}\\
         &=&\frac{1}{\cosh\delta}\exp\left(-\frac{x^2}{\alpha^2}-\frac{\xi^2}{\beta^2}\right).\nonumber
\end{eqnarray}

In the rest of this appendix, $A$ will always stand for operator $\boldsymbol{P}_{2\delta}^{(\gamma)}$, 
and $\lambda_k,\,k=0,1,\ldots,$ for its eigenvalues $\rho^{2k+1}\in(0,1)$. The proof of the subsequent 
Th.~\ref{TheoremAppendixB} follows the argument in \cite{JZ} (cf. also \cite{Jung}, \cite{FeiNo}), although 
the Szeg\H{o} theorems in \cite{JZ}, \cite{FeiNo} are inadequate for our purposes.
\begin{lemma}\label{LemmaAppendixB} For any polynomial $G_N(x,z)=\sum_{n=1}^N c_n(x) z^n$ 
with bounded variable coefficients $c_n(x)\in\mathbb{R}$, $x\in(1,\infty),$ it holds
\begin{equation}
  \sum_{k=0}^\infty G_N(\alpha\beta,\lambda_k)\doteq\frac{1}{2\pi}
         \int\!\!\!\int_{\mathbb{R}^2}G_N\left(\alpha\beta,\sigma_A(x,\xi)\right)\dxdxi.  
\end{equation}
\end{lemma}
\begin{proof}[Proof:] First, by (\ref{EXP}), for any $f\in L^2(\mathbb{R})$ it holds that
\[G_N(\alpha\beta,A)f=\sum_{k=0}^\infty G_N(\alpha\beta,\lambda_k)\langle f,
                                                              D_\gamma \psi_k \rangle D_\gamma \psi_k.
\]
Hence, operator $B=G_N(\alpha\beta,A)$ has the trace
\begin{equation}
  \mathrm{tr}\,B=\sum_{k=0}^\infty G_N(\alpha\beta,\lambda_k).\label{trace_1}
\end{equation}

Second, we use the key observation \cite[\emph{trace rule} (0.4)]{JZ} to obtain (here and thereafter, double 
integrals extend over $\mathbb{R}^2$)
\[
  \mathrm{tr}\,B=\frac{1}{2\pi}\int\!\!\!\int\sigma_B(x,\xi)\dxdxi,
\]
where $\sigma_B(x,\xi)$ is the Weyl symbol of operator $B$. By linearity of the Weyl correspondence, 
$\sigma_B(x,\xi)$ has the expansion 
\[
   \sigma_B(x,\xi)=\sum_{n=1}^N c_n(\alpha\beta)\sigma_{A^n}(x,\xi).
\]
Since for any $\gamma>0$ held constant the family of operators 
$\{\boldsymbol{P}_\delta^{(\gamma)};\delta>0\}$ forms a semigroup with respect to $\delta$ (see 
\cite{Ham2004}), it follows that $A^n=\boldsymbol{P}_{2n\delta}^{(\gamma)}$. In Eq. (\ref{Weyl_symbol}), 
replace operator $A$ by $A^n$ and $\delta$ by $n\delta$. Because of 
$\tanh(n\delta)=(n\tanh\delta)(1+o(1))$ we then obtain
\begin{eqnarray*}
  \sigma_{A^n}(x,\xi)&=&\frac{1}{\cosh(n\delta)}e^{-\tanh(n\delta)(\gamma^{-2}x^2+\gamma^2\xi^2)}\\
        &=&(1+o(1))\left(\sigma_A(x,\xi)\right)^n
	      \exp\left[-o(1)\left(\frac{x^2}{\alpha^2}+\frac{\xi^2}{\beta^2}\right)\right],
\end{eqnarray*}
where the Landau symbol $o(1)$ stands for various quantities vanishing as $\delta\rightarrow0$ (or 
$\alpha\beta\rightarrow\infty$). We now estimate
\begin{eqnarray}
  \mathrm{tr}\,B&=&\frac{1}{2\pi}\int\!\!\!\int \sigma_B(x,\xi)\dxdxi\nonumber\\
    &=&\left[\frac{1}{2\pi}\int\!\!\!\int G_N\left(\alpha\beta,\sigma_A(\alpha x,\beta\xi)\right)\,
                                                        dx\,d\xi +\epsilon\right]\alpha\beta\nonumber\\
    &=&\frac{1}{2\pi}\int\!\!\!\int G_N(\alpha\beta,\sigma_A(x,\xi))\dxdxi+\epsilon\,\alpha\beta,
                                                                                        \label{trace_2}
\end{eqnarray}
where $\epsilon\rightarrow 0$ as $\alpha\beta\rightarrow\infty$. Eq. (\ref{trace_2}) in combination 
with Eq. (\ref{trace_1}) concludes the proof.
\end{proof}
\begin{theorem}[Szeg\H{o} Theorem]\label{TheoremAppendixB} Let $g:[0,\Delta]\rightarrow\mathbb{R}$, 
$\Delta\in(0,\infty)$, be a continuous function such that $\lim_{x\rightarrow 0+}g(x)/x$ exists. For 
any functions $a,\,b:(1,\infty)\rightarrow\mathbb{R}$, where $a(x)$ is bounded and $b(x)\in[0,\Delta]$, 
define the function $G(x,z)=a(x)g(b(x)z),\,(x,z)\in(1,\infty)\times[0,1]$. Then it holds
\begin{equation}
  \sum_{k=0}^\infty G(\alpha\beta,\lambda_k)\doteq\frac{1}{2\pi}
          \int\!\!\!\int_{\mathbb{R}^2}G\left(\alpha\beta,\sigma_A(x,\xi)\right)\dxdxi.\label{Szego}
\end{equation}
\end{theorem}
\begin{proof}[Proof:]
The function $f(x)=g(x)/x,\,x\in(0,\Delta],$ has a continuous extension $F(x)$ onto 
the compact interval $[0,\Delta]$. By virtue of the Weiertstrass approximation theorem, for any 
$n\in\mathbb{N}$ there exists a polynomial $F_{N_n-1}(x)$ of some degree $N_n-1$ such that 
$|F(x)-F_{N_n-1}(x)|\le\epsilon_n=\frac{1}{n}$ for all $x\in [0,\Delta]$. Consequently, the polynomial 
$g_{N_n}(x)=xF_{N_n-1}(x)$ of degree $N_n$ satisfies the inequality
\begin{equation}
  |g(x)-g_{N_n}(x)|\le \epsilon_n x,\,x\in[0,\Delta]. \label{WAS_ineq}
\end{equation}

Define the polynomial with variable coefficients $G_{N_n}(x,z)=a(x)g_{N_n}(b(x)z)$. We now show that
\begin{equation}
  (\alpha\beta)^{-1}\sum_{k=0}^\infty G_{N_n}(\alpha\beta,\lambda_k)\rightarrow
        (\alpha\beta)^{-1}\sum_{k=0}^\infty G(\alpha\beta,\lambda_k) \label{first_arrow}
\end{equation}
and
\begin{eqnarray}
\lefteqn{\frac{(\alpha\beta)^{-1}}{2\pi}\int\!\!\!\int_{\mathbb{R}^2}G_{N_n}\left(\alpha\beta,
                                                        \sigma_A(x,\xi)\right)\dxdxi}\nonumber\\
        &\rightarrow&\frac{(\alpha\beta)^{-1}}{2\pi}\int\!\!\!\int_{\mathbb{R}^2}G\left(\alpha\beta,
	                                     \sigma_A(x,\xi)\right)\dxdxi, \label{second_arrow}
\end{eqnarray}
as $n\rightarrow\infty$, uniformly for $\alpha\beta\in(1,\infty)$.

\textit{Proof of (\ref{first_arrow}):} By Ineq. (\ref{WAS_ineq}) we get
\begin{eqnarray*}
|\sum_{k=0}^\infty G(\alpha\beta,\lambda_k)-\sum_{k=0}^\infty G_{N_n}(\alpha\beta,\lambda_k)|
                &\le&\sum_{k=0}^\infty|G(\alpha\beta,\lambda_k)-G_{N_n}(\alpha\beta,\lambda_k)|\\
                &\le& M\epsilon_n\Delta\sum_{k=0}^\infty \lambda_k,
\end{eqnarray*}
where $M=\sup\{|a(x)|;x>1\}<\infty$ and $\sum_{k=0}^\infty\lambda_k=\rho/(1-\rho^2)\le\alpha\beta/2$ for 
$\alpha\beta>1$. After devision of the last inequality by $\alpha\beta$, convergence in 
(\ref{first_arrow}) follows as claimed.

\textit{Proof of (\ref{second_arrow}):} Similarly,
\begin{eqnarray*}
\lefteqn{|\int\!\!\!\int G\left(\alpha\beta,\sigma_A(x,\xi)\right)\dxdxi
     -\int\!\!\!\int G_{N_n}\left(\alpha\beta,\sigma_A(x,\xi)\right)\dxdxi|}\\
       &\le&\int\!\!\!\int|G\left(\alpha\beta,\sigma_A(x,\xi)\right)
                                        -G_{N_n}\left(\alpha\beta,\sigma_A(x,\xi)\right)|\dxdxi\\
     &\le& M\epsilon_n\Delta\int\!\!\!\int\sigma_A(x,\xi)\dxdxi.
\end{eqnarray*}
Since $(2\pi)^{-1}\int\!\!\!\int\sigma_A(x,\xi)\dxdxi=\rho/(1-\rho^2)$, we arrive at the same 
conclusion as before.

Now, choose an arbitrarily large number $n\in\mathbb{N}$, substitute function $G$ in Eq.~(\ref{Szego}) 
by the polynomial $G_{N_n}$ and devide both sides of that equation by $\alpha\beta$. Then, 
by reason of Lem.~\ref{LemmaAppendixB} and uniform convergence in (\ref{first_arrow}) and 
(\ref{second_arrow}) with respect to $\alpha\beta\in(1,\infty)$, the theorem follows.
\end{proof}

\textit{Proof of Th.~\ref{form_1a}:}
Define 
\[
  \ln_+ x=\left\{\!\!\begin{array}{cl}\max\{0,\ln x\} & \mbox{if }x>0,\\
                                                0 & \mbox{if }x=0.
                     \end{array}\right.
\]
Because of Eq. (\ref{C1}) we have, recalling that $\sigma^2$ is dependent on $\alpha\beta$,
\begin{eqnarray*}
  C(S)&=&\sum_{k=0}^\infty\frac{1}{2}\ln_+\left(\frac{\sigma^2(\alpha\beta)}{\theta^2}
                                                                                     \lambda_k\right)\\
      &=&\sum_{k=0}^\infty a(\alpha\beta)g(b(\alpha\beta)\lambda_k),
\end{eqnarray*}
where $a(\alpha\beta)=1$, $b(\alpha\beta)=\sigma^2(\alpha\beta)/\theta^2$, 
$g(x)=\frac{1}{2}\ln_+x,x\in[0,\Delta]$, and $\Delta$ is chosen so that $b(\alpha\beta)\le\Delta<\infty$ 
when $\alpha\beta$ is large enough (the latter choice is possible since $\sigma^2(\alpha\beta)$ is 
finitely upper bounded as $\alpha\beta\rightarrow\infty$). Without loss of generality, we assume 
$b(\alpha\beta)\in[0,\Delta]$ for all $\alpha\beta\in(1,\infty)$. Then, by 
Th.~\ref{TheoremAppendixB} it follows that
\begin{eqnarray*}
  C(S)&\doteq&\frac{1}{2\pi}\int\!\!\!\int_{\mathbb{R}^2}
       \frac{1}{2}\ln_+\left(\frac{\sigma^2(\alpha\beta)}{\theta^2}\sigma_A(x,\xi)\right)\dxdxi\\
     &=&\frac{1}{2\pi}\int\!\!\!\int\frac{1}{2}
 \ln\left[1+\frac{\left(\frac{\sigma^2(\alpha\beta)}{2\pi}-N(x,\xi)\right)^+}{N(x,\xi)}\right]\dxdxi,
\end{eqnarray*}
where $N(x,\xi)=\frac{\theta^2}{2\pi}(\sigma_A(x,\xi))^{-1}$. Next, rewrite Eq. (\ref{def_sigma}) as
\[
   S=\sum_{k=0}^\infty\sigma^2(\alpha\beta)\left(1
                            -\frac{1}{\frac{\sigma^2(\alpha\beta)}{\theta^2}\lambda_k}\right)^+.
\]
Put $a(\alpha\beta)=\sigma^2(\alpha\beta)$, $b(\alpha\beta)=\sigma^2(\alpha\beta)/\theta^2$ and define
\[
    g(x)=\left\{\!\!\begin{array}{cl}\left(1-\frac{1}{x}\right)^+ & \mbox{if }x>0,\\
                                                                0 & \mbox{if }x=0.
                     \end{array}\right.
\] 
Without loss of generality, we assume that $a(\alpha\beta)$ is bounded for all 
$\alpha\beta\in(1,\infty)$. So, $b(\alpha\beta)\in[0,\Delta]$ where 
$\Delta=\sup\{a(\alpha\beta)/\theta^2;\alpha\beta>1\}<\infty$. Then, by Th.~\ref{TheoremAppendixB} it 
follows that
\begin{eqnarray*}
   S&\doteq&\frac{1}{2\pi}\int\!\!\!\int\sigma^2(\alpha\beta)\left(1-
      \frac{1}{\frac{\sigma^2(\alpha\beta)}{\theta^2}\sigma_A(x,\xi)}\right)^+\dxdxi\\
    &=&\int\!\!\!\int \left(\frac{\sigma^2(\alpha\beta)}{2\pi}-N(x,\xi)\right)^+\dxdxi.
\end{eqnarray*}
Finally, replacement of $\frac{\sigma^2(\alpha\beta)}{2\pi}$ by the parameter $\nu$ completes the proof.
\hfill $\Box$

\textit{Proof of Th.~\ref{R(D)_param}:} For any $\theta\in(0,\sigma]$ held constant 
define the distortion $D$ by Eq. (\ref{def_theta}) or, equivalently, by
\[
  D=\sum_{k=0}^\infty\theta^2\min\left\{1,\frac{\sigma^2}{\theta^2}\lambda_k\right\}.
\]
Since $D=\sum_{k=0}^\infty a(\alpha\beta)g(b(\alpha\beta)\lambda_k)$, where 
$a(\alpha\beta)=\theta^2$, $b(\alpha\beta)=\sigma^2/\theta^2$, $g(x)=\min\{1,x\}$ for $x\in[0,\Delta]$, 
$\Delta=\sigma^2/\theta^2$, it follows by Th.~\ref{TheoremAppendixB} that
\begin{eqnarray}
   D&\doteq&\frac{1}{2\pi}\int\!\!\!\int\theta^2\min\left\{1,\frac{\sigma^2}{\theta^2}
                                             \sigma_A(x,\xi)\right\}\dxdxi\label{Ddot_AppendixB}\\
   &=&\int\!\!\!\int\min\left\{\frac{\theta^2}{2\pi},
                                  \boldsymbol{\Phi}(t,\omega)\right\}\dt\domega, \nonumber
\end{eqnarray}
where $\boldsymbol{\Phi}(t,\omega)=\frac{\sigma^2}{2\pi}\,\sigma_A(t,\omega)$ is the WVS (\ref{WVS}). 
Next, rewrite Eq. (\ref{R1}) as
\[
  R=\sum_{k=0}^\infty\frac{1}{2}\ln_+\left(\frac{\sigma^2}{\theta^2}\lambda_k\right).
\]
Taking $a(\alpha\beta)=1$, $b(\alpha\beta)=\sigma^2/\theta^2$, 
$g(x)=\frac{1}{2}\ln_+x,x\in[0,\Delta]$, $\Delta$ chosen as before, we infer by 
Th.~\ref{TheoremAppendixB} that
\begin{eqnarray}
  R&\doteq&\frac{1}{2\pi}\int\!\!\!\int_{\mathbb{R}^2}
       \frac{1}{2}\ln_+\left(\frac{\sigma^2}{\theta^2}\sigma_A(x,\xi)\right)\dxdxi
                                                                              \label{Rdot_AppendixB}\\
      &=&\frac{1}{2\pi}\int\!\!\!\int\frac{1}{2}
       \ln_+\left[\frac{\boldsymbol{\Phi}(t,\omega)}{\frac{\theta^2}{2\pi}}\right]\dt\domega.
                                                                                             \nonumber
\end{eqnarray}

Finally, replacement of $\frac{\theta^2}{2\pi}$ by the parameter $\lambda$ completes the proof.
\hfill $\Box$
\end{appendix}

\end{document}